# The Universal Arrow of Time II: Quantum mechanics case.

## Kupervasser Oleg

## Abstract

The given paper is natural continuation of our previous paper [1]. We have illustrated earlier, that in the classical Hamilton mechanics for an overwhelming majority of real chaotic macroscopic systems there is alignment of their thermodynamic time arrows because of their small interaction. This fact and impossibility to observe entropy decrease at introspection explain the second low of thermodynamics. In a quantum mechanics the situation even is little bit easier - all closed systems of finite volume are periodic or nearly periodic. The proof in a quantum mechanics is in many respects similar to the proof in the classical Hamilton mechanics - it also uses small interaction between subsystems and impossibility to observe entropy decrease at introspection. However there are special cases which were not in the classical mechanics. In these cases a one microstate corresponds to a set of possible macrostates (more precisely their quantum superposition). Consideration this property with using decoherence theory and taking into account of thermodynamic time arrows introduces new outcomes in quantum mechanics. It allows to resolve the basic paradoxes of a quantum mechanics: (a) to explain paradox of wave packet reduction at measurings when the observer is included in system (introspection) (paradox of the Schrodinger cat); (b) to explain unobservability of superposition of macroscopic states by the external observer in real experiments (paradox of Wigner's friend); (c) to prove the full equivalence of multiworld and Copenhagen interpretations of quantum mechanics; (d) to explain deviations from the exponential law at decay of particles and pass from one energy level on another (paradox of a kettle which never will begin to boil).

## 1. Introduction

First of all, it is necessary to note, that in our paper if other is not stipulated, the full system is in the closed finite volume, contains a finite number of particles and is isolated from environment. These are principal requirements of the entropy increasing law which we consider. Full system is described by quantum mechanics laws.

In our previous paper [1] we considered alignment of thermodynamic time arrows in the classical Hamilton mechanics leading to the entropy increasing law. Here we wish to consider a quantum case. A reason of alignment of thermodynamic time arrows in a quantum mechanics is the same as in the classical mechanics. It is «*entangling*» and «*decoherence*» [2-3, 17, 24-27] - small interaction between real chaotic macroscopic systems or real chaotic macroscopic system in an unstable state and a quantum microsystem (measuring process in a quantum mechanics).

Use of phenomenon of alignment of thermodynamic time arrows for a quantum mechanics for the analysis of widely known paradoxes of a quantum mechanics gives their full and consistent resolution. All these paradoxes are caused by *experimental* unobservability for *real macroscopic* bodies of such purely quantum phenomena predicted by a quantum mechanics, as (a) superposition of macrostates for the Copenhagen interpretation or (b) presence of many worlds in case of multiworld interpretation.

Really, quantum mechanics has the principal difference from classical one - if in classical mechanics one microstate corresponds to only one macrostate, but for quantum mechanics one microstate (*a pure* state characterized by a wave function) can correspond to a set of macrostates. (Or, otherwise, the microsate is superposition of microstates corresponding to the *different* macrostates.) Such situation is not possible in the classical mechanics! Such state can not be considered as mixed state, i.e. ensemble of this several macrostates with corresponding probabilities. Evolution of superposition and the mixed states are different. This difference is related to presence *of the interference* terms for superposition (or *quantum correlations* of the worlds for multiworld interpretation). Though for macroscopic bodies this difference is very small, but, nevertheless, it exists. What disturbs to observe experimentally this difference? It is the same reasons that disturbs to observe entropy decreasing because of alignment of thermodynamic time arrows!

Really, more the detailed analysis below, shows, that the experimental manifestations of an interference (quantum correlations) are considerably only at entropy decrease. This process is not observable *in principle* if the observer *is included* into observable system (*introspection*). Thus, entropy decrease is very difficultly observable if the observer is not included in the observed macrosystem, because of alignment of thermodynamic time arrows of observable system and the observer/environment during decoherence. Almost full isolation of a macrosystem from environment / the observer is necessary between observations.

Small manifestations of the interference (quantum correlations) at entropy increase can not be observed at introspection *in principle* (at introspection the full observation is impossible - only macroparameters can be measured exactly, the full measuring is impossible). They are very difficultly observable for a case of the external observer because of decoherence with the observer/environment.

## 2. Qualitative consideration of the problem.

The reason of alignment of thermodynamic time arrows in a quantum mechanics, as well as in the classical mechanics, is small interaction between real chaotic macroscopic systems. This well studied appearance carrying a title «decoherence» [2-3, 17, 24-27]. Its result is not only widely known «entangling» states of systems, but also alignment of thermodynamic time arrows. (The direction of thermodynamic time arrow is defined by a direction of the entropy increase.) The reason of alignment of thermodynamic time arrows is the same, as in the classical Hamilton mechanics - instability of processes with opposite time arrows with respect to small perturbations. These perturbations exist between the observer/environment and observed system (decoherence).

Similar arguments in the case of quantum mechanics have been given in Maccone's paper [4]. However there he formulated, that the similar logic is applicable only in a quantum mechanics. The incorrectness of this conclusion has been shown in our previous papers [1, 5]. The other objection has been formulated in the paper [6]. There are considered small systems with strong fluctuations. Alignment of thermodynamic time arrows does not exist for such small systems. It must be mentioned that both Maccone's replay to this objection and the subsequent paper of objection authors [7] do not explain the true reason of described disagreement. The real solution is very simple. More specifically, the entropy increase law, the concept of thermodynamic time arrows and their alignment are applicable only to nonequilibrium *macroscopic* objects. Violation of these laws for microscopic systems with strong fluctuations is widely known fact. Nevertheless, though the objection [6] is trivial physically, but it is interesting from purely mathematical point of view. It gives good mathematical criterion for macroscopicity of chaotic quantum systems.

The situation in quantum mechanics even is easier, than in classical one: chaotic quantum systems are almost-periodic systems. Their chaotic character is defined by the fact that the energies (eigenvalues of a Hamiltonian, «frequencies» of energy modes) are distributed over the random law [8].

It is often possible to meet the statement that the behavior of quantum chaotic systems differs very strongly from the behavior of classical ones. It, however, the strong mistake related to deep misunderstanding physicists of these systems. Really, quantum chaotic systems are almost-periodic whereas classical chaotic systems are characterized by the random law for Poincare's returns times. Thermodynamic time arrows of the observer and the observable system have the same direction. Therefore the observer is capable to carry out observation (or introspection) only on finite time intervals when its time arrow exists (i.e. its state is far from thermodynamic equilibrium) and *does not change* its direction. On such *finite* times (that the observer is capable to carry out observation during this time) the behavior of chaotic quantum systems has the same character, as for classical quantum systems.

Decoherence results in pass of observed systems from a pure state to mixed one, i.e. results in entropy increase. (Actually, one macrostate transforms to the set of microstates). On the other hand, Poincare's returns yield inverse result (i.e. «recoherence») and are related to the entropy decrease. Decoherence and correspondent alignment of thermodynamic time arrows of the observer and observable systems leads, thus, also to the syncs of moments when the systems pass from pure states to mixed states. Consequently, it makes impossible to observe experimentally the inverse process (i.e. «recoherence»).

Summarizing, consideration of alignment of thermodynamic time arrows in quantum mechanics is in many aspects similar to the consideration in the classical mechanics. However consideration of this property for the analysis of widely known paradoxes of a quantum mechanics gives their full and consistent resolution. These are following paradoxes: (a) to explain paradox of wave packet reduction at measurings when the observer is included in system (introspection) (paradox of the Schrodinger cat); (b) to explain unobservability of superposition of macroscopic states by the external observer in real experiments (paradox of Wigner's friend); (c) to prove the full equivalence of multiworld and Copenhagen interpretations of quantum mechanics; (d) to explain deviations from the exponential law at decay of particles and pass from one energy level on another (paradox of a kettle which never will begin to boil).

As already described above, in quantum mechanics the solution of the problem of alignment of thermodynamic time arrows is similar to classical mechanics. But there is one important exception. In the classical mechanics one microstate (a point in a phase space) corresponds to only one macrostate. In the quantum mechanics one microstate (wave function) can corresponds to the set of possible macrostates (quantum superposition of the wave functions corresponding to this macrostates). This situation appears in well-known paradox of "Schrodinger cat".

Multiworld Interpretation of a quantum mechanics is very popular currently. It states, that these different macrostates corresponds to the different worlds. These parallel worlds exist simultaneously and interfere (summing with each other). it is suggested as a solution of «Schrodinger cat» paradox.

But then the following question appears: Why we need to guess simultaneous existence of these worlds? Instead we can say: «System collapses in one of these macrostates with the probability defined by Bohr's rules. Why we need these mysterious parallel worlds? ». This point of view is named Copenhagen Interpretation.
The following objections are usually given:
1. We do not have any mechanisms describing the collapse in Copenhagen Interpretation.
2. We accept that wave functions it is something really existing.
3. These wave functions and their superposition satisfy to Schrodinger equations.
4. Multiworld interpretation follows automatically from 1 and 2.
5. Decoherence, which also is a consequence of Schrodinger equations, explains why we can see as a result only one of the worlds (with corresponding Bohr's probabilities).

But it is possible to object here: «Yes, we have no collapse mechanism. But we need not know it. We postulate such collapse. Moreover, we at all do not want to know this mechanism. Really, we are capable to describe and calculate any physical situation without this knowledge».

But such approach meets following difficulties:
1. We cannot specify or calculate *an exact* instant when there is this collapse. For macrobodies it is possible to specify only very narrow, but, nevertheless, a finite interval of time on which this collapse happens.
2. For macrobodies there is enough clear split between the worlds (because of decoherence), but it is never full. Always there is *small* "overlapping" between the worlds (the interference terms, quantum correlations of the worlds) even for macrobodies. Decoherence, described above, resolves a problem only partially. It "separates" macroworlds not completely, leaving their small "overlapping".
3. There are specific models of collapse (so-called, GRW theory [16]). They can be verified experimentally. Such experiments did not give while any proof of existence of such collapse. They give only boundaries on parameters for such models (in the case that it is really true), defined by accuracy of experiment.

But it is possible to object again:

1. Yes, there is a problem to define exact collapse times. But precisely the same problem exists also in multiworld interpretation - in which instant the observer sees, in what of the possible worlds he has appeared?
2. The problem of "overlapping" of the worlds also exists in the multiworld interpretation. Really, the observer sees in some instant only a one world. He can tell nothing about existence or not existence of other parallel worlds. So all predictions of the future (based on the Bohr's rules) he can conclude only on knowledge «his» world. But because of "overlapping" of the worlds (really very small) some effects appear which can not be based on his predictions. It means that quantum mechanics can not give even the exact *probability* prediction.
3. It is possible to add one more uncertainty that exists in both interpretations. Let, for example that a superposition of two macrostates exists: «a live cat» and «a dead cat». Why the world splits (or collapses) on such two states? Why is a pair: («a live cat» - «a dead cat»), («a live cat» + «a dead cat») not relevant?

Three problems described above lead *to uncertainty* of the predictions done from quantum mechanics. It can not be found even within frameworks based on Bohr's rules. This uncertainty is very small for macrobodies, but it exists. It exists for *all* interpretations, only masking and changing its form.
   Majority of interpretations try to overcome these problems. Actually different interpretation only "masks" the uncertainty problem, not solving it.

4.   All told above about GRW theories is true. There is no necessity to use it instead of quantum mechanics. However it is not correct for Copenhagen Interpretation. The Copenhagen Interpretation reminds GRW very much, but one important feature is very *differs* from GRW. The Copenhagen Interpretation postulates the collapse only for *one* final observer. It does not demand the collapse from the *rest* macroobjects and observers. Physical experiment is described from a point of view of this final observer. The final "observer" is not some person possessing mysterious "consciousness". It is some standard macroscopic object. It is far from its state of thermodynamic equilibrium. The final observer is last in the chain of observers and macrobodies. Direction of his thermodynamic time arrows is chosen as "positive" direction. It is similarly to our previous paper [1]. This constrain on collapse leads to serious consequence which does not appear in GRW. Namely, the existence of the collapse in GRW can be verified experimentally, but in Copenhagen Interpretation the existence of the collapse can not be proved or disproved even *in principle.* Let us demonstrate it. We will consider thought experiments which allow to verify existence of the collapse predicted in GRW. Further we will demonstrate that these experiments can not be used for verification of the collapse in the Copenhagen Interpretation.

a. Quantum mechanics, as well as classical, predicts Poincare's returns. And, unlike classical chaotic systems, the returns happen periodically or almost periodically. But because of the collapse in GRW such returns are impossible and can not observed experimentally. I.e. this fact can be used for experimental verification.

b. Quantum mechanics is reversible. At a reversion of evolution the system must return to an initial state. However, the collapse results in irreversibility. This fact also can be verified experimentally.
c. We can observe experimentally the small effects related to the small quantum correlations which exist even after decoherence. In GRW this small effects disappear.

Suppose that we want to verify the collapse of the final observer in the Copenhagen Interpretation. Hence, we must include the observer to observable system. I.e. here there is *introspection*. We will demonstrate, that it is not impossible to verify existence (or not existence) of the collapse in Copenhagen Interpretation by the methods described above:

a. Suppose, the observer waits for the return predicted by a quantum mechanics. But the observer is included to the system. I.e. at Poincare's return, he will return to his initial state together with the full system. Hence, his memory about his past will be erased. So the observer will not be possible to compare of an initial and finite state. It makes the verification of the existence (or not existence) of the observer's collapse experimentally impossible.
b. The same reasons, as in item (a.), make impossible the experimental verification of the returns caused by the reversion of system evolution.
c. For observation of the small effects (quantum correlation macrostates), the measuring split-hair accuracy is necessary. But, as the observer is included into observed system (introspection) it is not possible to make full measurement of such system. (Figuratively speaking, the observer uses some "ink" to describe the full system state. But the "ink" is also part of the full system during intersection. So the "ink" must describe also itself!) Such system can be described by macroparameters only. It makes impossible experimental observation and calculation of the small effects of the quantum correlations.

As a matter of fact, first two items (a., b.) are related to a following fact which took place also in classical mechanics [1]. Decoherence (decomposition on macrostates) leads to the entropy increase (one macrostate replaces on the full set of possible macrostates). On the other hand, observation of the return (i.e. recoherence) is related to the entropy decrease. The observer is capable to carry out experimentally introspection only on finite time intervals when it has time arrow (i.e. a state far from thermodynamic equilibrium), and it *does not change* its direction. Thus, impossibility experimentally to distinguish the Copenhagen and Multiworld Interpretations it is closely related to the entropy increase law and the thermodynamic arrow of time.

Everything told above, makes impossible the experimental verification of the difference between the Copenhagen and Multiworld Interpretation. It makes their equivalent. Such statements about indistinguishability of these interpretations meet in the literature. However there, where this fact is usually stated, it is usually referred to impossibility to make such verification only practically for macrobodies (FAPP - for all practical purposes). The understanding of its *principal* impossibility is absent. This incorrect understanding is a basis for erroneous deduction about «exclusiveness» of Multiworld Interpretation. We will demonstrate the clearest example [9]:

**"Proponents of the MWI might argue that, in fact, the burden of an experimental proof lies on the opponents of the MWI, because it is they who claim that there is new physics beyond the well tested Schrodinger equation."**
"Despite the name "interpretation", the MWI is a variant of quantum theory that is different from others. Experimentally, the difference is relative to collapse theories. It seems that there is no experiment distinguishing the MWI from other no-collapse theories such as Bohmian mechanics or other variants of MWI. The collapse leads to effects that are, in principle, observable; these effects do not exist if the MWI is the correct theory. To observe the collapse we would need a super technology, which allows "undoing" a quantum experiment, including a reversal of the detection process by macroscopic devices. See Lockwood 1989 (p. 223), Vaidman 1998 (p. 257), and other proposals in Deutsch 1986. These proposals are all for gedanken experiments that cannot be performed with current or any foreseen future technology. Indeed, in these experiments an interference of different worlds has to be observed. Worlds are different when at least one macroscopic object is in macroscopically distinguishable states. Thus, what is needed is an interference experiment with a macroscopic body. Today there are interference experiments with larger and larger objects (e.g., fullerene molecules $C_{60}$), but these objects are still not large enough to be considered "macroscopic". Such experiments can only refine the constraints on the boundary where the collapse might take place. A decisive experiment should involve the interference of states which differ in a macroscopic number of degrees of freedom: an impossible task for today's technology"

The correct proof of principal experimental unverifiability of collapse in Copenhagen Interpretation, as far as we know, meets only in this and previous papers [10-13]. It is possible to term it as the "Godel" theorem of impossibility for a quantum mechanics. Both its statement and its proof method really remind «the Godel theorem of incompleteness».

We here concentrate on this problem so much because of the following reasons. At first, the impossibility experimentally to distinguish the Copenhagen and Multiworld Interpretations is closely related to the entropy increase law and the thermodynamic arrow of time. Secondly, it is too much people sincerely, but wrongly believes, that Multiworld Interpretation (or other less fashionable Interpretations) completely solves all problems of quantum mechanics. Uncertainty already described above is such problem of quantum mechanics. It means that quantum mechanics using Bohr's rules have small uncertainty connected to small quantum correlation of the observer. How they are solved actually? These results can be concluded from the fact that the specified uncertainty exist in *ideal* dynamics over abstract coordinate time. In *observable* dynamics over the observer's time arrow it misses and is not observed experimentally in principle.

> 1) Introspection. The same reasons already described above which do not allow to verify the collapse experimentally will not allow to discover experimentally the uncertainty specified in item 1 (an exact instant of the collapse) and item 2 (quantum correlations). So it is senseless to discuss it.
> 2) External observation:

a. If this observation does not perturb observable system then the collapse of the system and, hence, and indeterminacies [specified in item 1 (an exact instant of the collapse) and item 2 (quantum correlations)] does not arise. So quantum mechanics can be verified precisely experimentally. Such unpertrubative observation is possible for macrobodies only theoretically. The necessary condition is a known initial state (pure or mixed) (Appendix A).
b. the observed system is open. It means that there is a small interaction between observable system and the observer/environment. This small interaction masks uncertainty (specified in points 1 and 2) and makes impossible its experimental observation.

Here it is necessary to return to the uncertainty described in item 3. The majority of real observations correspond to two cases: the introspection cases (when the full description is impossible in principle) or the open system (perturbed with uncontrollable small external noise from the observer/environment). How to describe such open or incomplete systems? It becomes by input *of macroparameters* of system. Real *observable* dynamics of such parameters is possible for a wide class of systems. It does not include unobservable in realities «the parallel worlds», entropy reduction, quantum superposition of macrostates and other exotic, possible only in *ideal* dynamics. Observable dynamics is considered with respect to the thermodynamic time arrow of the real macroscopic nonequilibrium observer, weakly interacting with observable system and an environment (decoherence). Ideal dynamics is considered with respect to abstract, coordinate time. The problem of the pass from ideal to real dynamics is successfully solved in other papers [14-15, 17-18]. The select of macrovariables is ambiguous, but also is not arbitrary. Macrovariables should be chosen so that at entropy increase random small external noise did not influence considerably their dynamics. Such macrovariables exist and are named pointer states [3, 17]. Presence of the selected states is a result of interaction locality in the real world. It means that the close particles interact stronger than far particles. If the force of interaction was defined, for example, by closeness of momentums the principal states would be absolutely others. So, the property of a locality is untrue over distances comparable with wave length. So radiowaves have *field* pointer states, strongly differing from *particles* pointer states. A situation here described, completely equivalent to [1] where were considered "appropriate" macrostates for classical mechanics.

What can be an example observable dynamics for quantum systems? These are described above GRW theories. To understand it we will return to the Copenhagen Interpretation. We can choose for "the final observer" in the Copenhagen Interpretation different nonequilibrium macrobodies. Theoretically, thus the collapse will see differently for such different observers. This appearance is named «paradox of Wigner's friend». This appearance of ambiguity of the collapse in the Copenhagen Interpretation can be named «Quantum solipsism». It is made by analogy to similar philosophical doctrine. This problem can be resolved similarly to the paper [1]. The entropies of all weakly interacting macrobodies increase or decrease synchronously, because of alignment of thermodynamic time arrows. The collapse corresponds to entropy increase (one macrostate replaces on a set of possible macrostates). Hence, small interaction (decoherence) between macrobodies yields not only alignment of thermodynamic time arrows, but also sync of all moments of «collapse» for different observers. It makes «Quantum solipsism» for macrobodies though theoretically possible, but it is the extremely difficult realizable in practice. So this resolution of «Quantum solipsism» by the collapses differs from Copenhagen Interpretation where the observer's collapse cannot be prevented even theoretically. Discussed above GRW theories are, thus, for quantum mechanics description of the real *observable* dynamics of macrobodies (FAPP dynamics). It throws out effects not observed in reality. It is, for example, non synchronism in the macrobodies collapses moments and entropy decrease predicted by *ideal* dynamics.

Good illustration of the above described connection observed and ideal dynamicses is «the paradox of a kettle which never will begin to boil». It is related in quantum mechanics to a deviation from the exponential law of particles decay (or a pass from one energy level on another). Exponential character of such law is very important - the relative rate of decay does not depend on an instant. It means that the decaying particle has no "age". In a quantum mechanics, however, on small times the law *of ideal* dynamics of decay strongly differs from the exponential one. So when the number of measurements of a decaying particle state for finite time interval increases, the particle in limit of infinite number of measurements does not decays at all!

Let we observe a macrosystem consisting of major number of decaying particles. Here it is necessary to note, that particle decay happens under laws of ideal dynamics only between measurements. Measurements strongly influence dynamics of system as we described above. To transfer to *the observable* dynamics featured above, we should decrease perturbing influence of observation strongly. It is reached by increasing interval between observations. It must be comparable with a mean lifetime of unperturbed particles. For such large intervals of time, we get real observable dynamics of decay. It is featured by an exponential curve, and the mean lifetime does not depend on a concrete interval between measurements. Thus, the exponential decay is a law of observable dynamics, not ideal dynamics of particles. (The same reason explains absence of Poincare's returns for this system.)

## 3. The quantitative consideration of the problem.

### 3.1 Definition of the basic concepts.

1) In classical mechanics a microstate is a point in a phase space. In quantum mechanics it corresponds to a wave function $\psi$ (a pure state), and trajectories is wave function evolution in time. In classical mechanics the macrostate corresponds to a density distribution function in a phase space. In quantum mechanics it corresponds to a density matrix $\rho$. The density matrix form depends on the chosen basis of orthonormal wave functions. If $\rho\rho \neq \rho$ then it is in mixed state.

2) The equations of motion for the density matrix $\rho$ have the form:

$$i\frac{\partial \rho_N}{\partial t} = L\rho_N,$$

where $L$ is the linear operator:

$L\rho = H\rho - \rho H = [H, \rho]$

and H is the energy operator of the system,

N - number of particles

3) If $A$ is the operator of a certain observable, then the average value of the observable can be found as follows:

$<A> = tr\, A\, \rho$

4) If the observation is introspection the full observation to is impossible. In case of external observation because of small interaction with the observer and instabilities of a observable chaotic system the full exposition also is senseless. Therefore introducing a finite set M of *macrovariables* is necessary:

$A_{set} = \{A_1, A_2, ..., A_M\}$,

Where $M << N$

These macrovariables are known with finite small errors:

$\Delta A_i << A_i,\ 1 \leq i \leq M$

This set of macrovariables corresponds to a macrostate with a density matrix $\rho_{set}$.

All microstates answering to requirements

$\{|<A_1> - A_1| \leq \Delta A_1,\ |<A_2> - A_2| \leq \Delta A_2,\ ...,\ |<A_M> - A_M| \leq \Delta A_M\}$

are assume to have equal probabilities.

*To thermodynamic equilibrium* corresponds to a macrostate $\rho_E$. It corresponds to a set of the microstates, satisfying to a requirement

$|<E> - E| \leq \Delta E\ (\Delta E << E)$,

where $E$ is the full system energy.

All these microstates are assumed to have equal probabilities.

5) In quantum mechanics *ensemble entropy* is defined via density matrix **[15]**:

$S = -k\, tr\, (\rho\, ln\, \rho)$,

where *tr* stands for matrix trace.

Entropy defined in such a way does not change in the course of reversible evolution:

$$\frac{\partial S}{\partial t} = 0$$

*6) Macroscopic entropy* is defined as follows:

a) For current $\rho$ we find all corresponding sets of macrovariables

$$\begin{cases} A_{set}^{(1)} = \{A_1^{(1)}, A_2^{(1)}..., A_M^{(1)}\}\ \Delta A_i^{(1)} << A_i^{(1)}, 1 \leq i \leq M \\ \vdots \\ A_{set}^{(L)} = \{A_1^{(L)}, A_2^{(L)}..., A_M^{(L)}\}\ \Delta A_i^{(L)} << A_i^{(L)}, 1 \leq i \leq M \end{cases}$$

b) We find a matrix $\rho_{set}$ for which all microstates, corresponding to the specified set of macroparameters, have equal probabilities

c) Macroscopic entropy $S = -k\, tr\, (\rho_{set}\, ln\, \rho_{set})$

Unlike ensemble entropy macroscopic entropy (macroentropy) is not constant and can both to increase, and to decrease in time. For given energy $E \pm \Delta E$ it reaches its maximum for thermodynamic equilibrium. The direction of the macroentropy increase defines a direction of a *thermodynamic arrow of time* for the system.

7) Like classical case the interaction locality results in fact what not all macrostates are appropriate. They should be chosen so that small noise did not influence essentially system evolution for entropy increase process. Such states are well investigated in quantum mechanics and named *pointer states* [3, 17]. Quantum superposition of such states is unstable with respect to small noise. So such superposition is not, accordingly, pointer state. For macrosystems, close to the equilibrium, pointer states are usually correspondent to Hamiltonian eigenfunctions.

8) *Coarsened* value of $\rho$ ($\rho_{coar}$) should be used to obtain changing entropy similarly to changing macroscopic entropy. We will enumerate ways to achieve it:

   a) We define a set of pointer states and we project a density matrix $\rho$ on this set. I.e. (a) we note a density matrix $\rho$ in representation of these pointer states (b) we throw out nondiagonal terms of $\rho$ and obtain $\rho_{coar}$. So entropy:
   $S = -k\, tr\, (\rho_{coar}\, ln\, \rho_{coar})$

   b) We divide the system into some interacting subsystems (for example: the observer, the observable system and the environment). Then we define the full entropy as the sum of the entropies of these subsystems:
   $S = S_{ob} + S_{ob\_sys} + S_{env}$

## 3.2 Effect of a weak coupling

### 3.2.1 Small external perturbation.

We can put our macrosystem of finite volume inside of an infinite volume system ("environment", "reservoir") with some temperature. (This reservoir can be also vacuum with zero temperature.) We will suppose that this reservoir is in thermodynamic equilibrium, has the same temperature as a temperature of the finite system in equilibrium and weakly interacts with our finite system. Then it is possible to use the quantum version of "new dynamics", developed by Prigogine [14] for such infinite systems. Dynamics of our finite system with reservoir will be the same as its *observable dynamics* without reservoir with respect to its thermodynamic time arrow. Such description has sense only during finite time. It is time when the its thermodynamic time arrow exists (i.e. the system is not in equilibrium) and does not change its direction.

### 3.2.2 Alignment of thermodynamic time arrows at interaction of macrosystems (the observer and the observable system).

It ought to be noted, that here our job is much easier, than in case of the classical mechanics. This is due to the fact that the quantitative theory of small interaction between quantum systems (*decoherence, entangling*) is well developed field [2-3, 17, 24-27]. We will not repeat these conclusions here. We will give only short results:

**(a)** Suppose that we have in some instant two macrosystems. One of them or both these macrosystems are in their quantum superposition of pointer states. The decoherence theory [2-3, 17, 24-27] states, that small interaction between macrosystems very fast (decoherence time is much less than relaxation time to equilibrium) transforms such system into the mixed state. So the quantum superposition disappears. Such process of vanishing of quantum superposition of pointer states corresponds to the entropy increase. It follows from Poincare's theorem that the system (in coordinate time) should return to its initial state. There should be an inverse process of recoherence. But it will happen in both systems synchronously. It means, that any system can see only the decoherence and the entropy increase with respect to its thermodynamical time arrow. It means that both processes decoherence and time arrows will be synchronous in interacting subsystems. It is especially worthy of note that we consider here a case *of macroscopic* systems. For small systems where large fluctuations of parameters are possible, similar alignment of thermodynamic time arrows and the instances of «collapses" for subsystems is not observed [6-7].

**(b)** Now suppose that all macroscopic subsystems are in them pointer states. In the decoherence theory it is shown, that in the presence of small noise between its macroscopic subsystems the behavior of a quantum system is completely equivalent and is indistinguishable from behavior of the correspondent classical system [2-3, 17 , 24-27]. Thus, the analysis of alignment of thermodynamic time arrows is completely equivalent to the analysis made in paper [1].

**(c)** It is worthy of explanation which meaning the words "classical system" have here. It means, that in the theory there do not exist specific mathematical features of quantum theory. It is, for example, such features as not commuting observables, quantum superposition of pointer states. At that, these "classical theories" can be very exotic, include Plank's constant and are not reduced to laws of the known mechanics of macrobodies or waves.

Superconductivity, superfluidity, radiation of absolute black body, the superposition of currents in Friedman's experiment [19] is often named "quantum effects". They are really quantum in the sense that their equations of motion include Planck constant. But they are perfectly featured over macroscale by a mathematical apparatus of usual classical theories: or the theory of *classical* field (as pointer states), or the theory of *classical* particles (as pointer states). From this point of view, they are not quantum, but classical. In quantum theory featured objects simultaneously are both particles and probability waves.

It is worthy of note that in classical limit at room temperatures the quantum mechanics *of large* particles gives the theory *of classical* particles as pointer states (electron beams, for example). On the other hand *light in weight* particles give the *classical* field as pointer states (radiowaves). And these theories do not include Planck constant.

However, at high temperatures when radiation achieves high frequencies, light quanta are featured by the theory *of classical* particles as pointer states. They give, for example, a spectrum of absolute black body on high frequencies. Though this spectrum includes Planck's constant its dynamics of pointer states (particles) will be classical. For deriving this spectrum the quantum mechanics formalism is not necessary (Planck has derived this spectrum, nothing knowing about quantum physics mathematical apparatus).

Vice versa at low temperatures the particles start to be featured by *classical* fields as pointer states (superfluidity or superconductivity phenomena). For example, superconductivity is featured by *classical wave* of "order parameter". And though the equations, featuring this field, include Planck constant, but the equations correspond to mathematical apparatus of the *classical* field theory. These waves can be summed (superpose) with each other, similarly quantum. But quadrate of their amplitude does not define probability density. It defines density of Cooper pair. Such wave cannot collapse at measurement, like a probability quantum wave [20].

For quantum-mechanical states of bosons at low temperatures pointer states are *classical* fields, and at high temperatures it's *classical* particles. The word "classical" is understood as a mathematical apparatus of the observable dynamics featuring their behavior, but not presence or lack of Planck constant in their equations of motion.

What happens in the intermediate states between classical fields and classical particles? It is, for example light in an optical wave guide ($L >> \lambda >> \lambda_{ultraviolet}$), $L_{opt}$ - the characteristic size of the macrosystem (the optical wave guide) (Appendix B), $\lambda$ - light-wave length, $\lambda_{ultraviolet}$- ultra-violet boundary of light). At use of macroscales and macrovariables, and taking into account small noise from the observer both descriptions («classical waves» and «classical beam of particles») are identical. They are equivalent and can be used as pointer states. The equivalent situation arises for a superconductor case where the role of particles or waves play elemental "excitation" in gas of Cooper pairs.

Let's carry out the simple calculation illustrating above.

Let E be energy of particle; k -Boltzmann constant, T-temperature, p - momentum, $\Delta p$ - momentum uncertainty, $\lambda$ - particle wave length, $\omega$-frequency, $\Delta x$ - a coordinate uncertainty; $\hbar$ - Planck constant. We will consider the "gas" of such particles which is in a cavity, filled with some material with distance between atoms a. a <<L, L - the characteristic size of the cavity. In vacuum a ~ $(L^3/N)^{1/3}$, N-number of particles in the cavity. c - light velocity (let suppose for simplicity that refraction index in the cavity is close to 1).

**1)** Firstly, let consider light in weight particles which at room temperature have a velocity, close to light velocity c.
E~pc; E~kT; p ~$\Delta p$; $\lambda$ ~$\Delta x$; $\Delta p \Delta x$ ~ $\hbar$; $\omega$ =E/$\hbar$
From here
$\hbar$ ~ $\Delta p \Delta x$ ~ p$\lambda$ ~ kT$\lambda$/c => $\lambda$ ~ $\hbar$c/kT
Requirement *of classical* field approach with frequency $\omega$ ~c/$\lambda$:
L <$\lambda$ or L ~ $\lambda$. From here L <$\hbar$c/kT or L~$\hbar$c/kT
Requirement of approach *of classical* relativistic particles with E~$\hbar$c/$\lambda$ and p=E/c:
L>> $\lambda$. From here L>> $\hbar$c/kT

**2)** Secondary, let consider heavy particles *bosons* which at room temperature have a velocity v<<c
p ~ $(Em)^{1/2}$; E~kT; p ~ $\Delta p$; $\lambda$ ~ $\Delta x$; $\Delta p \Delta x$ ~ $\hbar$; $\omega$ =E/$\hbar$
From here
$\hbar$ ~ $\Delta p \Delta x$ ~ p$\lambda$ ~ $(kTm)^{1/2}\lambda$ => $\lambda$~$\hbar$ /$(kTm)^{1/2}$
Requirement *of classical* field approach with frequency $\omega$ =$p^2$/(m $\hbar$):
L <$\lambda$ or L ~ $\lambda$. From here L <$\hbar$ /$(kTm)^{1/2}$ or ~$\hbar$/$(kTm)^{1/2}$
Requirement of approach *of classical* particles with energy E=$p^2$/(2m) and momentum p=mv:
L>> $\lambda$. From here L>> $\hbar$/$(kTm)^{1/2}$

**3)** Let consider now heavy particles *fermions* which at room temperature have a velocity v <<c
p ~ $(Em)^{1/2}$; E~kT; p ~$\Delta p$; $\Delta p \Delta x$ ~ $\hbar$
$\Delta x \leq \lambda$ and $\lambda \leq a$ is a requirement of Pauli's principle for fermions. They cannot appear in the same state, so they are distributed in "boxs" with size a.
From here
$\hbar$ ~ $\Delta p \Delta x \leq$ p$\lambda$ ~ $(kTm)^{1/2}\lambda$ => a $\geq \lambda \geq$ $\hbar$/$(kTm)^{1/2}$
T$\geq T_F$ = $\hbar^2$/($a^2$km) - Fermi's temperature when fermion gas transfers in a ground state and expression E~kT becomes untrue.
At T <$T_F$: E~$E_F$=k$T_F$; $\lambda$~ $\hbar$/$(E_F m)^{1/2}$~ a
Requirement of *classical* field approach:
$L < \lambda$ or $L \sim \lambda$. But it is impossible because $L >> a \geq \lambda$

At $T \geq T_F$ approach of *classical* particles in quality pointer states with energy $E = {p^2}/{2m}$ and momentum $p = mv$ is correct.

At $T \leq T_F$ approach of classical particles in quality pointer states, prisoners in «boxes» with size a, with energy $E \sim E_F$ and momentum $p \sim (E_F m)^{1/2}$ is correct.

At $T \sim T_F$ we observe dynamics of "excitations" in the degenerated Fermi gas which is featured by particles or waves as pointer states for these "excitations".

To create in experiment the paradox of "Schrodinger cat", the quantum superposition of the pointer states is necessary, instead of superposition of classical waves. Therefore superposition of classical waves of "order parameter" or light waves is not related in any way to this paradox and does not illustrate it.

So, for example, experiment Friedman [19] states a superposition of opposite currents. But the superposition is itself pointer state for this case. This pointer state is classical, not *quantum* superposition of pointer states as untruly it is usually declared. Really, the state of bosons system (Cooper pairs) is featured at such low temperature *by a classical* wave as it was demonstrated above. These waves of "order parameter" are pointer states. They differ from pointer states of a high-temperature current of *classical particles* having a well-defined direction of a motion. The superposition observed in Friedman experiment, is not capable to collapse to quantum-mechanical sense: Its quadrate features not probability, but density of Cooper pairs [20]. It is no more surprising and no more "quantum", than usual superposition of electromagnetic modes in the closed resonator where spectrum of modes is discrete also. The only difference is that "order parameter" wave equations for pointer states include ℏ. It is the only reason to use concept "quantum" for this case.

### 3.3 Resolution of Loshmidt and Poincare paradoxes within the framework of quantum mechanics.

The state of quantum chaotic system in the closed cavity with finite volume is featured by a set of energy modes $u_k(r_1, ..., r_N)$ with spectrum $E_k$ distributed under *the random law* [8]. Let's write the expression for wave functions of a noninteracting pair of such systems:

$$\psi^{(1)}(\boldsymbol{r}_1,...,\boldsymbol{r}_N,t) = \sum_k u_k(\boldsymbol{r}_1,...,\boldsymbol{r}_N) e^{-\frac{iE_k^{(1)}}{\hbar}t}$$

$$\psi^{(2)}(\boldsymbol{r}_1,...,\boldsymbol{r}_L,t) = \sum_l v_l(\boldsymbol{r}_1,...,\boldsymbol{r}_L) e^{-\frac{iE_l^{(2)}}{\hbar}t}$$

The united equation is following:

$$\psi^{(1)}(\boldsymbol{r}_1,...,\boldsymbol{r}_N,\boldsymbol{r}_1,...,\boldsymbol{r}_L,t) = \psi^{(1)}(\boldsymbol{r}_1,...,\boldsymbol{r}_N,t)\psi^{(2)}(\boldsymbol{r}_1,...,\boldsymbol{r}_L,t) =$$

$$\sum_k \sum_l u_k(\boldsymbol{r}_1,...,\boldsymbol{r}_N) v_l(\boldsymbol{r}_1,...,\boldsymbol{r}_L) e^{-\frac{i(E_k^{(1)}+E_l^{(2)})}{\hbar}t}$$

At presence small interactions between the systems
$$\psi^{(1)}(\boldsymbol{r}_1,...,\boldsymbol{r}_N,\boldsymbol{r}_1,...,\boldsymbol{r}_L,t) =$$

$$\sum_k \sum_l f_{kl}(\boldsymbol{r}_1,...,\boldsymbol{r}_N,\boldsymbol{r}_1,...,\boldsymbol{r}_L) e^{-\frac{iE_{kl}}{\hbar}t},$$

where $E_{kl} = E_k^{(1)} + E_l^{(2)} + \Omega_{kl}$, $\Omega_{kl}$-generally a set of random variables, $f_{kl}$, $u_k$, $v_l$ are eigenfunctions of corresponding Hamiltonians.

Received solutions are almost-periodic functions. Received period of return defines Poincare's period. Period of Poincare's return of full system is generally more periods of both subsystems.

For resolution of Poincare and Loshmidt paradoxes (returns in these paradoxes contradict to entropy increase law) we will consider now three cases

*1) Introspection*: At introspection the time arrow is always directed over entropy growth so the observer is capable to see only entropy growth with respect to this time arrow. Besides, return to an initial state erases memory about past. It does not allow the observer to detect entropy reduction. Thus, reduction of entropy and returns happen only with respect to coordinate time. But any experiment is possible with only with respect to time arrow of the observer. With respect to coordinate time entropy reduction and returns cannot be experimentally observed [1, 10-13].

2) *External observation with small interaction* between macrosystems: Small interaction results in alignment of the thermodynamic time arrows of the observer and observed systems. Accordingly, all arguments that are relevant for introspection again become relevant for this case.

3) For very hardly realizable *experiment with unperturbative observation* (Appendix A) macroentropy reduction can really be observed. However, it is worthy of note, that in the real world "entropy costs" on the experimental organization of such unperturbative observations will exceed considerably this entropy decrease. Indeed, the observable system needs to be isolated very strongly from environment noise.

In classical systems the period of Poincare's return is the random variable strongly depending on an initial state. In quantum chaotic systems the period is well defined and does not depend considerably on an initial state. However, this real difference in behavior of quantum and classical systems is not observed experimentally even in absence of any explicit constraint on experiment time. Indeed, any real physical experiment has a duration that is much smaller than Poincare's period of macrobodies. Physical experiments are possible only during time while the thermodynamic time arrow exists (i.e. the system is not in a state of thermodynamic equilibrium) and does not change the direction.

## 3.4 Decoherence for process of measurement

### 3.4.1 Reduction of system at measurement [22-23].

Let's consider a situation when a measuring device was at the beginning in state $|\alpha_0\rangle$, and the object was in superposition of states $|\psi\rangle = \sum c_i |\psi_i\rangle$, where $|\psi_i\rangle$ are experiment eigenstates. The initial statistical operator is given by expression

$\rho_0 = |\psi\rangle |\alpha_0\rangle\langle\alpha_0| \langle\psi|$ (1)

The partial track of this operator which is equal to statistical operator of the system, including only the object, looks like

$tr_A(\rho_0) = \sum_n \langle\varphi_n|\rho_0|\varphi_n\rangle$

where $|\varphi_n\rangle$ - any complete set of device eigenstates. Thus,

$tr_A(\rho_0) = \sum |\psi\rangle \langle\varphi_n|\alpha_0\rangle\langle\alpha_0|\varphi_n\rangle\langle\psi| = |\psi\rangle\langle\psi|$, (2)

Where the relation $\sum |\varphi_n\rangle\langle\varphi_n| = 1$ and normalization condition for $|\alpha_0\rangle$ are used. We have statistical operator correspondent to object state $|\psi\rangle$. After measuring there is a correlation between device and object states, so the state of full system including device and object is featured by a state vector

$|\Psi\rangle = \sum c_i e^{i\theta_i} |\psi_i\rangle |\alpha_i\rangle$. (3)

And the statistical operator is given by expression

$\rho = |\Psi\rangle\langle\Psi| = \sum c_i c_j^* e^{i(\theta_i - \theta_j)} |\psi_i\rangle|\alpha_i\rangle\langle\alpha_j|\langle\psi_j|$. (4)

The partial track of this operator is equal to

$tr_A(\rho) = \sum_n \langle\varphi_n| \rho |\varphi_n\rangle =$
$= \sum_{(ij)} c_i c_j^* e^{i(\theta_i - \theta_j)} |\psi_i\rangle \{\sum_n \langle\varphi_n|\alpha_i\rangle\langle\alpha_j|\varphi_n\rangle\}\langle\psi_j| =$
$= \sum_{(ij)} c_i c_j^* \delta_{ij} |\psi_i\rangle\langle\psi_j|$ (5)

(Since various states $|\alpha_i\rangle$ of device are orthogonal each other); thus,

$tr_A(\rho) = \sum |c_i|^2 |\psi_i\rangle\langle\psi_i|$. (6)

We have obtained statistical operator including only the object, featuring probabilities $|c_i|^2$ for object states $|\psi_i\rangle$. So, we come to formulation of the following theorem.

**Theorem 1** (about measuring). If two systems *S* and *A* interact in such a manner that to each state $|\psi_i\rangle$ systems *S* there corresponds a certain state $|\alpha_i\rangle$ of systems *A* the statistical operator $tr_A(\rho)$ over full systems *(S* and *A)* reproduces wave packet reduction for measuring, yielded over system *S,* which before measuring was in a state $|\psi\rangle = \sum_i c_i |\psi_i\rangle$.

Suppose that some subsystem is in mixed state but the full system including this subsystem is in pure state. Such mixed state is named *as improper mixed state*.

### 3.4.2 The theorem about decoherence at interaction with the macroscopic device. [21, 22]

Let's consider now that the device is a macroscopic system. It means that each distinguishable configuration of the device (for example, position of its arrow) is not a pure quantum state. It states nothing about a state of each separate arrow molecule. Thus, in the above-stated reasoning the initial state of the device $|\alpha_0\rangle$ should be described by some statistical distribution on microscopic quantum states $|\alpha_{0,s}\rangle$; the initial statistical operator is not given by expression (1), and is equal

$$\rho_0 = \sum_s p_s |\psi\rangle |\alpha_{0,s}\rangle \langle\alpha_{0,s}|\langle\psi|. \tag{7}$$

Each state of the device $|\alpha_{0,s}\rangle$ will interact with each object eigenstate $|\psi_i\rangle$. So, it will be transformed to some other state $|\alpha_{i,s}\rangle$. It is one of the quantum states of set with macroscopic description correspondent to arrow in position i; more precisely we have the formula

$$e^{iH\tau/\hbar}(|\psi\rangle |\alpha_{0,s}\rangle) = e^{i\theta_{i,s}} |\psi\rangle |\alpha_{i,s}\rangle. \tag{8}$$

Let's pay attention at appearance of phase factor depending on index *s*. Differences of energies for quantum states $|\alpha_{0,s}\rangle$ should have such values that phases $\theta_{i,s}$ (mod $2\pi$) after time $\tau$ would be randomly distributed between 0 and $2\pi$.

From formulas (7) and (8) follows that at $|\psi\rangle = \sum_i c_i |\psi_i\rangle$ the statistical operator after measuring will be given by following expression:

$$\rho = \sum_{(s,i,j)} p_s c_i c_j^* e^{i(\theta_{i,s} - \theta_{j,s})} |\psi_i\rangle |\alpha_{i,s}\rangle \langle\alpha_{j,s}|\langle\psi_j| \tag{9}$$

As from (9) the same result (6) can be concluding. So we see that the statistical operator (9) reproduces an operation of reduction applied to given object. It also practically reproduces an operation of reduction applied to device only ("practically" in the sense that it is a question about "macroscopic" observable variable). Such observable variable does not distinguish the different quantum states of the device corresponding to the same macroscopic description, i.e. matrix elements of this observable variable correspondent to states $|\psi_i\rangle |\alpha_{i,s}\rangle$ and $|\psi_j\rangle |\alpha_{j,s}\rangle$ do not depend on *r* and *s*. Average value of such macroscopic observable variable *A* is equal to

$$\text{tr}(\rho A) = \sum_{(s,i,j)} p_s c_i c_j^* e^{i(\theta_{i,s} - \theta_{j,s})} \langle\alpha_{j,s}|\langle\psi_j|A|\psi_i\rangle |\alpha_{i,s}\rangle =$$
$$= \sum_{(i,j)} c_i c_j^* a_{i,j} \sum_s p_s e^{i(\theta_{i,s} - \theta_{j,s})} \tag{10}$$

As phases $\theta_{i,s}$ are distributed randomly, the sum over s are zero at i≠j; hence,

$$\text{tr}(\rho A) = \sum |c_i|^2 a_{ii} = \text{tr}(\rho' A). \tag{11}$$

where

$$\rho' = \sum |c_i|^2 p_s |\psi_i\rangle |\alpha_{i,s}\rangle \langle\alpha_{j,s}|\langle\psi_j| \tag{12}$$

We obtain statistical operator which reproduces operation of reduction on the device. If the device arrow is observed in position i, the device state for some s will be $|\alpha_{i,s}\rangle$. The probability to find state $|\alpha_{i,s}\rangle$ is equal to probability of that before measuring its state was $|\alpha_{i,s}\rangle$. Thus, we come to the following theorem.

**Theorem 2. About decoherence of the macroscopic device**. Suppose that the quantum system interacts with the macroscopic device in such a manner that there is a chaotic distribution of states phases of the device. Suppose that $\rho$ is a statistical operator of the device after the measuring, calculated with the help of Schrodinger equations, and $\rho'$ is the statistical operator obtained as a result of reduction application to operator $\rho$. Then it is impossible to yield such experiment with the macroscopic device which would register difference between $\rho$ and $\rho'$. It is the so-called Daneri-Loinger-Prosperi theorem **[21]**.

For a wide class of devices it is proved that the chaotic character in distribution of phases formulated in the theorem 2 really takes place if the device is macroscopic and chaotic with unstable initial state. Indeed, randomness of phase appears from randomness of energies (eigenvalues of Hamiltonian) in quantum chaotic systems [8].

It is worth to note that though Eq. (12) is relevant with a split-hair accuracy it is only assumption with respect to (9). There from it is often concluded that the given above proof is FAPP. It means that it is only difficult to measure quantum correlations practically. Actually they continue to exist. Hence, *in principle* they can be measured. It is, however, absolutely untruly. Really, from Poincare's theorem about returns follows that the system will not remain in the mixed state (12), and should return to the initial state (7). It is the result of the very small corrections (quantum correlation) which are not included to (12). Nevertheless, the system featured here | $α_{i,s}$ ⟩ corresponds *to the introspection* case, and consequently, it is not capable to observe experimentally these returns *in principle (*as it was shown above in resolution of Poincare and Loshmidt paradoxes). Hence, effects of these small corrections exist only on paper in the coordinate time of ideal dynamics, but it cannot be observed *experimentally* with respect to thermodynamic time arrow of observable dynamics of the macroscopic device. So, we can conclude that Daneri-Loinger-Prosperi theorem actually results in a complete resolution (not only FAPP!) of the reduction paradox *in principle*. It proves impossibility to distinguish *experimentally* the complete and incomplete reduction.

The logic produced here strongly reminds Maccone's paper [4]. It is not surprising. Indeed, the pass from (7) to (12) corresponds to increasing of microstates number and entropy growth. And the pass from (12) in (7) corresponds to the entropy decrease. Accordingly, our statement about experimental unobservability to remainder quantum correlation is equivalent to the statement about unobservability of the entropy decrease. And it is proved by the similar methods, as in [4]. The objection [6] was made against this paper. Unfortunately, Maccone could not give the reasonable replay [28] to this objection. Here we will try to do it ourselves.

Let's define here necessary conditions.
Suppose A is our device, and C is the measured quantum system.
The first value, the mutual entropy *S (A: C)* is the coarsened entropy of ensemble (received by separation on two subsystems) excluding the ensemble entropy. As the second excluding term is constant, so *S (A: C)* describes well the behavior of macroentropy in time:

$S (A: C) = S (\rho_A) + S (\rho_C) - S (\rho_{AC})$,

Where $S = - tr (\rho \ln \rho)$,
The second value *I (A: C)* is the classical mutual information. It defines which maximum information about measured system ($F_j$) we can receive from indication of instrument ($E_i$). The more correlation exists between systems, the more information about measured system we can receive:

$I (A: C) = max_{E_i \otimes F_j} H (E_i:F_j)$, where

$H (E_i: F_j) = \Sigma_{ij} P_{ij} \log P_{ij} - \Sigma_i p_i \log p_i - \Sigma_j q_j \log q_j$

and $P_{ij} = Tr [E_i \otimes F_j \rho_{AC}]$, $p_i = \Sigma_j P_{ij}$ and $q_j = \Sigma_i P_{ij}$
given POVMs (Positive Operator Valued Measure) $E_i$ and $F_j$ for A and C respectively.
Maccone [4] proves an inequality
$S (A: C) \geq I (A: C)$ (13)
He concludes from it that entropy decrease results in reduction of the information (memory) about the system A+C and C.
But (13) contains an inequality. Correspondingly in [6] an example of the quantum system of three qubits is supplied. For this system the mutual entropy decrease is accompanied by mutual information increases. It does not contradict to (13) because mutual entropy is only up boundary for mutual information there.
Let's look what happens in our case of the macroscopic device and the measured quantum system
Before measurement (7)
$S (A: C) = - \Sigma_s p_s \log p_s + 0 + \Sigma_s p_s \log p_s = 0$

$E_i$-corresponds to the set $|\alpha_{0,s}\rangle$, $F_j$ - $|\psi\rangle$

$I(A:C) = -\sum_s p_s \log p_s + 0 + \sum_s p_s \log p_s = 0 = S(A:C)$

In the end of measurement from (12)

$S(A:C) = -\sum_i |c_i|^2 \log |c_i|^2 - \sum_{s,i} |c_i|^2 p_s \log |c_i|^2 p_s + \sum_{s,i} |c_i|^2 p_s \log |c_i|^2 p_s = -\sum_i |c_i|^2 \log |c_i|^2$

$E_i$-corresponds to the set $|\alpha_{i,s}\rangle$, $F_j$ - $|\psi_j\rangle$

$I(A:C) = -\sum_i |c_i|^2 \log |c_i|^2 - \sum_{s,i} |c_i|^2 p_s \log |c_i|^2 p_s + \sum_{s,i} |c_i|^2 p_s \log |c_i|^2 p_s =$
$-\sum_i |c_i|^2 \log |c_i|^2 = S(A:C)$

Thus, our case corresponds to

$I(A:C) = S(A:C)$                                            (14)

in (13). No problems exist for our case. It is not surprising – the equality case in (13) corresponds to macroscopic chaotic system. The system supplied by the objection [6] is not microscopic. It demonstrates the widely known fact that such *thermodynamic* concepts as the thermodynamic time arrows, the entropy increase and the measurement device concern to macroscopic chaotic systems. Both the paper [6] and the subsequent paper [7] describe not thermodynamic time arrows but, mainly, strongly fluctuating small systems. No thermodynamics is possible for such small systems as three cubits. The useful outcome of these papers is equality (14). It can be used as a *measure for macroscopicity* of chaotic quantum systems. On the other hand, the difference between mutual information and mutual entropy can be a criterion of fluctuations value.

   The paper of David Jennings, Terry Rudolph "Entanglement and the Thermodynamic Arrow of Time" is very interesting. But the *Thermodynamic* Arrow of Time is not applicable for microsystems. It is a nice paper about quantum fluctuation, but not a paper about *Thermodynamic* Arrow of Time. In the Abstract of the paper "Entanglement and the Thermodynamic Arrow of Time" the authors write: "We examine in detail the case of three qubits, and also propose some simple experimental demonstrations possible with small numbers of qubits." But no thermodynamics is possible for such a microsystem. D. Jennings and T. Rudolph (like Maccone) don't understand that category "*thermodynamic* arrow of time" is correct only for large macrosystems. Using these categories for small fluctuating systems has no physical sense. They also (like Maccone) use incorrect definition of macroscopic *thermodynamic* entropy. We also give (instead of Maccone) the correct reply to "Comment on "Quantum Solution to the Arrow-of-Time Dilemma"". The *correct* reply is that no contradictions (found in this Comment) appear for macroscopic systems. Only for a microscopic system such contradictions exist. But the concepts "the Thermodynamic Arrow of Time" and "the entropy growth law" is not relevant for such systems. We illustrate this fact by consideration of a quantum chaotic macrosystem and demonstrate that no contradiction (found by David Jennings, Terry Rudolph for a microscopic system) exists for this correct thermodynamical case. It must be mentioned that big size of a system (quantum or classic) is also not an enough condition for a system to be macroscopic. The macroscopic system (considered in Thermodynamics) must also be chaotic (quantum or classic) and has small chaotic interaction with its environment/observer resulting in decoherence (for quantum mechanics) or decorrelation (for classical mechanics). It should be also mentioned that thermodynamic-like terminology is widely and effectively used in quantum mechanics, quantum computers field, and information theory. The big number of the examples can be found in the references of Jennings's and Rudolph's paper. The other nice example is Shannon's entropy in information theory. But usually an author (using such a thermodynamic-like terminology) does not consider such a paper as analysis of classical Thermodynamics. Contrarily Jennings and Rudolph "disprove" the second law of Thermodynamics on the basis of the irrelevant microscopic system (in their Comment) and give (also in this Comment) the announcement of their next paper «Entanglement and the *Thermodynamic* Arrow of Time" as a correct consideration and a disproof of the second law.

## 4. Conclusion.

In the paper the analysis of thermodynamic time arrow in quantum mechanics is presented. It is in many aspects similar to classical case. The important difference of quantum systems from classical ones exists. One microstate in quantum mechanics can correspond not to one macrostate, but to a set of macrostate. It is named quantum superposition of macrostates. For this case considering thermodynamic time arrow by means of the decohernce theory give resolution of the quantum paradoxes. These paradoxes related to a wave packet reduction (collapse).

**Appendix A. Unpertrubative observation in the quantum and classical mechanics.**

It is often possible to meet a statement, that in the classical mechanics in principle it is possible always to organize unpertrubative observation**.** On the other hand in a quantum mechanics interaction of the observer with the observable system at measurement is inevitable. We will show that both these statements are generally untrue.

Let us first define the nonperturbative observation [10-11, 30-31] in QM. Suppose we have some QM system in a known initial state. This initial state can be either a result of some preparation (for example, an atom comes to the ground electronic state in vacuum after long time) or a result of a measurement experiment (QM system after measurement can have a well defined state corresponding to the eigenfunction of the measured variable). We can predict further evolution of the initial wave function. So *in principle* we can make further measurements choosing measured variables in such a way that one of the eigenfunctions of the current measured variable is a current wave function of the observed system. Such measuring process can allow us the continuous observation without any perturbation of the observed quantum system. This nonperturbative observation can be easily generalized for the case of a known *mixed* initial state. Really, in this case the measured variable at each instant should correspond to such set of eigenfunctions that the density matrix in representation of this set at the same instant would be diagonal.

For example, let us consider some quantum computer. It has some well-defined initial state. An observer that known this initial state can *in principle* make the nonperturbative observation of any intermediate state of the quantum computer.

It is especially worthy of note that such unpertrubative observation is possible only under condition of a known initial state. But, an observer that doesn't know the initial state can not make such observation, because he can not predict the intermediate state of the quantum computer.

Let's consider now classical mechanics. Suppose that a grain of sand lies on a cone vertex. The grain of sand has *infinitesimally small* radius. The system is in the Earth field of gravity. Then attempt to observe system even with *infinitesimal perturbation* will lead to a disbalance with the indefinite future through *a terminating* interval of time. Certainly, the reduced example is exotic - it corresponds to a singular potential and an infinitesimal object. Nevertheless, similar strongly labile systems are good classical analogues of quantum systems. Among them it is possible to search for analogies to quantum systems and quantum paradoxes. Having introduced a requirement, that classical measuring renders very small, but not zero perturbation on measured system, it is possible to lower requirements to a singularity of these systems.

Very often examples can be met of "purely quantum paradoxes", which do not ostensibly have analogy in the classical mechanics. One of them is Elitzur-Vaidman paradox **[**29**]** with a bomb which can be found without its explosion:

*Suppose that the wave function of one light quantum branches on two channels. In the end these channels the waves again unite, and there is an interference of the two waves of probability. A bomb inserted to the one from the two channels will destroy process of interference. Then it allows us to discover the bomb even for a case when the light quantum would not detonate it, having transited on other channel. (The light quantum is considered capable to detonate the bomb)*

Classical analogy of this situation is the following experiment of classical mechanics:

*In the one of the channels where there is no bomb, we throw in a macroscopic beam of many particles. In other channel where, maybe, there is the bomb, we will throw in simultaneously only one **infinitesimally easy** particle. Such particle is not capable to detonate the bomb. It will be throw out it back. If the bomb is not present, the particle will transit the channel. On an exit of this channel for the bomb we will arrange the cone featured above with the grain of sand with **infinitesimal** radius on the cone vortex. If our infinitesimally easy particle would throw down the grain of sand from the vertex it means, that the bomb is not present. If the grain of sand would remains on the vertex after an exit of particles beam from the second channel it means, that the bomb is.*

In the given example infinitesimally easy particle is analogue of an "imponderable" wave function of the light quantum. But the light quantum is sensitive to behavior of this "imponderable" wave function. As well the grain of sand with infinitesimal radius on the cone vertex is sensitive with respect to infinitesimally easy particle.

Summing up, it is possible to say, that the difference between quantum and classical systems is not so fundamental, as it is usually considered.

**Application B. Expansion on modes at arbitrary boundary conditions**.

Often there is a problem of description of radiation in a closed cavity filled by some substance. Usually it becomes by expansion of radiation on modes. These modes are a set of eigenfunctions of the wave equation for some cavity and for some boundary conditions. For example, it is a square cavity with periodical boundary conditions. Then the received radiation expansion is substituted to the wave equation for radiation. There the modes of the series are differentiated termwise. Thus, such radiation characteristic, as ω (k) is received. Here ω is frequency of a mode; **k** is a mode wave vector; $|k| = 2\pi/\lambda$; $\lambda$ is a mode wave length.

But here there is a purely mathematical problem. Suppose that the modes have been discovered for some shape of the cavity and for some boundary conditions. For termwise differentiability uniform convergence in all points of space is required. It is automatically true for any radiation with the same shape of a concavity and boundary conditions as modes. But for any other case it not true. Modes are the full orthogonal set and any radiation it is possible to present as superposition of such modes. But generally the series converges nonuniformly (the series converges badly near cavity boundaries) and can not be termwise differentiable. The problem of possible necessity using different modes for different boundary conditions is discussed in Peierls's book **[32].** However there is considered a case when some complete orthonormal set of modes exists for given boundary conditions. But a situation is possible that for such boundary condition no set of such modes are possible. Or the boundary conditions are not known, and only energy requirements on boundary are known. How can the problem be solved for such cases?

The point is that all perturbations in radiation are expending with a velocity which is not exceeding a light velocity in cavity v=c. It means, that any perturbation of initial conditions of radiation expands from a point x to a point $x_1$ only over finite time $(x-x_1)/c$. It means, that perturbations from walls will reach the cavity centre in time t=L/c, where L - the characteristic size of the cavity. Nonuniform convergence appears only near the cavity walls. So inside of the cavity far from walls the exact radiation field is almost precisely equal to the modes series during time L/c. Therefore this field has uniform convergence and can be termwise differentiable during time L/c.

To estimate correctly frequency of a mode ω (**k**) it is necessary, that its amplitude does not change essentially from walls perturbation over time t>>T. T=2π/ω(**k**) is time period of the mode. Therefrom we receive a requirement of cavity macroscopicity:

$2\pi/\omega \ll L/c$

or

$L \gg 2\pi(c/\omega)$

$\omega$ - correspondent to maximum of frequencies $\omega(\mathbf{k})$.

Let that this condition is fulfilled.

It means, that termwise differentiation of modes far from concavity walls can be made over timescales $t < 2\pi/\omega = L/c$.

On timescales $t > L/c$ the outcome cannot be correct. Here usually use the energy conservation law and the entropy increase law. By means of these laws slow evolution of amplitudes $A(t, r)$ and phases $\varphi(t, r)$ of modes can be received:

$E(t, \mathbf{r}) = \Sigma_i A_i(t, \mathbf{r}) \sin(\omega(\mathbf{k}_i) t + \mathbf{k}_i \mathbf{r} + \varphi_i(t, \mathbf{r}))$

For vacuum:

$\omega(\mathbf{k}) = c|\mathbf{k}|$
$L \gg \lambda$

# Acknowledgment

We thank Hrvoje Nikolic and Vinko Zlatic for discussions and debates which help very much during writing this paper.

# Универсальная стрела времени II: Случай квантовой механики.

Купервассер О.Ю.

## Аннотация.

Данная статья является естественным продолжением нашей предыдущей статьи [1]. Мы ранее проиллюстрировали, что в классической Гамильтоновой механике для подавляющего большинства реальных хаотических макроскопических систем происходит синхронизация собственных стрел времени вследствие их малого взаимодействия. Этот факт и невозможность наблюдать убывание энтропии при самонаблюдении объясняют второе начало термодинамики. В квантовой механике ситуация даже немного проще – все замкнутые системы конечного объема являются периодическими или почти периодическими. Доказательство в квантовой механике во многом аналогично доказательству в классической Гамильтоновой механике – оно использует учет малого взаимодействия между подсистемами и невозможность наблюдать убывание энтропии при самонаблюдении. Однако имеются особые случаи, которых не было в классической механике. В этих случаях одному микросотоянию соответствует несколько возможных макросотояний (точнее их квантовая суперпозиция). Рассмотрение этого свойства с использованием теории декогеренции и учета термодинамической стрелы времени привносит новые результаты в квантовую механику. Оно позволяет разрешить основные парадоксы квантовой механики: (а) объяснить парадокс редукции квантового пакета при измерениях, когда наблюдатель включен в систему (самонаблюдение) (парадокс Шредингеровского кота); (б) объяснить ненаблюдаемость суперпозиции макроскопических состояний внешним наблюдателем в реальных экспериментах (парадокс друга Вигнера); (в) доказать полную эквивалентность многомировой и Копенгагенской интерпретаций квантовой механики; (г) объяснить отклонения от экспоненциального закона при распаде частиц и переходах с одного энергетического уровня на другой (парадокс котелка, который никогда не закипит).

## 1. Введение

Прежде всего, следует отметить, что в нашей статье, если не оговорено иное, полная система находится в замкнутом ограниченном объеме, содержит конечное число частиц и изолирована от остальной части Вселенной. Это главные условия закона роста термодинамической энтропии, который мы будем обсуждать. Она также описывается законами квантовой механики.

В нашей предыдущей статье [1] мы рассматривали синхронизацию стрел времени в классической Гамильтоновой механике и вытекающее из него доказательство закона роста энтропии. Здесь мы хотим рассмотреть квантовый случай. Причиной синхронизации стрел времени в квантовой механике, как и в классической механике, являются «*перепутывание*» и «*декогеренция*» [2-3, 17, 24-27] - малое взаимодействие между реальными хаотическими макроскопическими системами или реальной хаотической макроскопической системой в неустойчивом состоянии и квантовой микросистемой (процесс измерения в квантовой механике).

Использование явления синхронизации стрел времени на квантовую механику для анализа широко известных парадоксов квантовой механики дает их полное и непротиворечивое разрешение. Все эти парадоксы связаны с *экспериментальной* ненаблюдаемостью для *реальных макроскопических* тел таких чисто квантовых явлений, предсказываемых квантовой механикой, как (а) суперпозиция состояний для Копенгагенской интерпретации или (б) наличие многих миров в случае многомировой интерпретации.

Действительно, квантовая механика обладает принципиальным отличием от классической – если для классической механики одному микросостоянию соответствует только одно макросотояние, то для квантовой механики одному микросостоянию (*чистое* состояние, описываемое волновой функцией) может соответствовать несколько макросотояний. (Или, иными словами, это микросотояние является суперпозицией микросостояний, соответствующих *разным* макросостояниям.) Ситуация не представимая в классической механике! Причем такое состояние не может рассматриваться как просто *смешанное* состояние, т.е. классический ансамбль нескольких макросостояний (точнее соответствующих им микросостояний, входящих в суперпозицию) с соответствующими вероятностями. Эволюция таких суперпозиций и смешанных состояний отличается. Это отличие связано с наличием *интерференционных* членов для суперпозиции (или *квантовых корреляций* миров для многомировой интерпретации). Хотя для макроскопических тел эти различия очень малы, но, тем не менее, они существуют. Что же мешает их экспериментально наблюдать? Те же причины, что препятствуют наблюдать уменьшение энтропии, вследствие синхронизации стрел времени!

Действительно, более подробный анализ, проводимый ниже, показывает, что экспериментальные проявления интерференции (квантовых корреляций) проявляются *значительно* лишь в момент убывания энтропии. Процесс этот не наблюдаем *в принципе*, если наблюдатель *входит* в наблюдаемую систему (*самонаблюдение*). При этом, он очень трудно наблюдаем (требует почти полной изоляции макросистемы от декогеренции окружения/самого наблюдателя) для макросистем, внешних по отношению к наблюдателю из-за синхронизации стрел времени наблюдаемой системы и наблюдателя/окружения при декогеренции.

Малые же проявления интерференции (квантовых корреляций) при росте энтропии также не могут наблюдаться при самонаблюдении *в принципе* (из-за принципиального ограничения в их точности – при самонаблюдении могут измеряться лишь макропараметры, полное измерение невозможно). Они же очень трудно наблюдаемы для случая внешнего наблюдателя из-за декогеренции с наблюдателем/окружением.

## 2. Качественное рассмотрение вопроса.

Причиной синхронизации стрел времени в квантовой механике, как и в классической механике, является малое взаимодействие между реальными хаотическими макроскопическими системами. Это хорошо изученное явление, носящее название «*декогеренции*» [2-3, 17, 24-27]. Ее результатом является не только широко известное «*перепутывание*» состояний систем, но и синхронизация их временных стрел. (Направление стрелы времени определяется направлением роста энтропии.) Причина такой синхронизации абсолютно та же, что и в классической Гамильтоновой механике – неустойчивость процессов с убыванием энтропии по отношению к малым возмущениям со стороны наблюдателя/окружения (декогеренция).

Похожие аргументы в случае квантовой механики были даны в работе Maccone [4]. Однако там он утверждал, что подобная логика применима только в квантовой механике. Ошибочность подобного взгляда была показана в наших предыдущих работах [1,5]. Кроме того было выдвинуто и другое возражение против него в работе [6]. Там рассматриваются небольшие, сильно флюктуирующие системы, в которых нарушается синхронизация стрел времени. Следует отметить, что как ответ на это возражение самого Maccone, так и последующая работа самих авторов возражения [7] не объясняют истинную причину замеченного несоответствия. Она же очень проста и заключается в том, что закон возрастания энтропии, само понятие термодинамической стрелы времени и их синхронизация – применимы только к неравновесным *макроскопически* объектам. Нарушение этих законов для микроскопических, сильно флюктуирующих систем – широко известный факт. Тем не менее, хотя само возражение [6] тривиально физически, но оно интересно с чисто математической точки зрения. Оно дает хороший математический *критерий макроскопичности* хаотических квантовых систем.

Ситуация в квантовой механике даже проще, чем в классической – здесь хаотическим квантовым системам соответствую почти периодические системы. Их хаотичность проявляется в том, что энергии, характеризующие собственные значения гамильтониана и определяющие «частоты» энергетических мод, распределены по случайному закону [8].

Часто можно встретить утверждение, что квантовые хаотические системы по своему поведению очень сильно отличаются от классических хаотических систем. Это, однако, сильное заблуждение, связанное с глубоким непониманием физики этих систем. Действительно, квантовые хаотические системы почти периодические, тогда как классические хаотические системы характеризуются случайным законом для времен возврата Пуанкаре. У наблюдателя и наблюдаемой системы стрелы времени синхронизированы. Поэтому наблюдатель способен экспериментально проводить наблюдение (или самонаблюдение) лишь на ограниченных промежутках времени, когда у него существует стрела времени (т.е. состояние далекое от термодинамического равновесия), и она *не меняет* свое направление. На таких *конечных* и экспериментально реально наблюдаемых временах поведение хаотичных квантовых систем носит тот же характер, что и для классических квантовых систем.

Декогеренция приводит к переходу наблюдаемых подсистем из чистого состояние в смешанное, т.е. приводит к росту энтропии. (На самом деле, ведь одно макросостояние заменяется на целый набор возможных макросостояний.) С другой стороны, возвраты Пуанкаре дают обратный результат (т.е. «рекогеренцию») и связаны с уменьшением энтропии. Декогеренция и сопутствующая ей синхронизация стрел времени наблюдателя и наблюдаемых подсистем приводит, таким образом, также к синхронизации моментов перехода из чистого состояние в смешанное всех этих подсистем и невозможности экспериментально наблюдать обратный процесс (т.е. «рекогеренцию»).

Подводя итог вышесказанного, рассмотрение явления синхронизации стрел времени в квантовой механике во многом аналогично рассмотрению в классической механике. Однако рассмотрение этого свойства для анализа широко известных парадоксов квантовой механики дает их полное и непротиворечивое разрешение. Это следующие парадоксы: (а) парадокс редукции квантового пакета при измерениях (парадокс Шредингеровского кота); (б) ненаблюдаемость суперпозиции макроскопических состояний (парадокс друга Вигнера); (в) *строгое* доказательство полной эквивалентности многомировой и Копенгагенской интерпретаций квантовой механики; (г) отклонения от экспоненциального закона при распаде частиц и переходах с одного энергетического уровня на другой (парадокс котелка, который никогда не закипит).

Как уже указывалось выше решение вопроса о синхронизации стрел времени в квантовой механике аналогично классической механике. Но имеется одно важное исключение. В классической механике одному микросостоянию (точке в фазовом пространстве) соответствует только одно макросостояние. В квантовой же механике одному микросостоянию (волновая функция) может соответствовать целый набор возможных макросотояний (квантовая суперпозиция волновых функций, соответствующих отдельным макросостояниям). Эта ситуация возникает в известном парадоксе «Шредингеровского кота».

Сейчас очень популярна Многомировая Интерпретация квантовой механики. Она утверждает, что этим отдельным макросостояниям соответствуют отдельные и одновременно существующие параллельно миры, которые интерферируют (складываются друг с другом). В этом видится разрешение парадокса «Шредингеровского кота».

Но тут возникает первый, на первый взгляд тривиальный вопрос: зачем нам нужно предполагать одновременное существование этих миров. Вместо этого мы можем просто сказать: «Система коллапсирует в одно из этих макросотояний с вероятностью, определяемой правилами Бора. Для чего нам нужны эти мистические параллельные миры?». Данному взгляду отвечает Копенгагенская Интерпретация.

На это обычно даются следующие возражения
1. У нас нет никаких механизмов, описывающих коллапс Копенгагенской Интерпретации.

2. Мы принимаем, что волновые функции это нечто реально существующее.
3. Эти волновые функции и их суперпозиция удовлетворяют уравнению Шредингера.
4. Из 1 и 2 автоматически следует многомировая интерпретация.
5. Декогеренция, которая также является следствием уравнения Шредингера, объясняет, почему мы можем в итоге видеть лишь один из миром (с соответствующей вероятностью Бора).

Но на это вполне можно возразить: «Да, у нас нет механизма коллапса. Но нам и не нужно его знать. Мы просто постулируем наличие такого коллапса. Более того, мы даже не хотим знать этот механизм, поскольку способны описать и рассчитать любую физическую ситуацию без этого знания».

Но такой подход встречает следующие трудности:
1. Мы не можем указать или рассчитать *точный* момент времени, когда происходит этот коллапс. Для макротел можно указать лишь очень узкий, но, тем не менее, конечный интервал времени, на котором этот коллапс происходит.
2. Для макротел существует достаточно четкое разделение между мирами (за счет декогеренции), но оно никогда не является полным. Всегда остается *небольшое* «перекрытие» между мирами (интерференционные члены, квантовые корреляции миров) даже для макротел. Декогеренция, описываемая выше, лишь частично решает проблему. Она не до конца «разъединяет» макромиры, оставляя это небольшое их «перекрытие».
3. Существуют специфические модели коллапса (так называемая, GRW теория [16]). Они могут быть проверены экспериментально. Такие эксперименты не дают пока никакого доказательства существования такого коллапса. Они дают лишь границы на параметры таких моделей (в том случае, если они всё-таки верны), определяемые точностью эксперимента.

На это можно снова возразить:

1. Да, есть проблема определить точное время коллапса. Но точно такая же проблема существует и в многомировой интерпретации – в какой именно момент времени наблюдатель видит, в каком из возможных миров он очутился?
2. Тоже касается и «перекрытия» миров – эта проблема присутствует и в многомировой интерпретации. Действительно, наблюдатель видит в некоторый момент времени только свой мир. Он ничего не может сказать о наличии и отсутствии других параллельных миров. Соответственно, все предсказания будущего (определяемые правилами Бора) могут делаться им на основе знания лишь «своего» мира. Но из-за «перекрытия» миров (пусть и малого) могут возникнуть эффекты, не укладывающиеся в эти предсказания. То есть, квантовая механика при ее таком последовательном использовании не способна давать даже точный *вероятностный* прогноз.
3. Можно добавить и еще одну неопределенность, присущую обеим интерпретациям. Пусть, к примеру, имеются два макросотояния – «живой кот» и «мертвый кот». Почему мир разделяется (или коллапсирует) именно на такие два состояния? Чем хуже, например, пара: («живой кот» - «мертвый кот»), («живой кот» + «мертвый кот»)?

Указанные выше три проблемы приводят к *неопределенности* в предсказаниях, делаемых квантовой механикой. Она не укладывается даже в вероятностные рамки, определяемые правилами Бора. Эта неопределенность очень мала для макротел, но она существует. Она присутствуют в *любых* интерпретациях, лишь маскируясь и меняя свою форму.

Невероятный поток интерпретаций связан именно с попыткой преодолеть эти проблемы. На самом деле разные интерпретация лишь по разному «маскируют» проблему неопределенности, не решая ее.

4. Все сказанное выше о GRW теории верно. Нет никакой необходимости заменять ею квантовую механику. Однако Копенгагенская Интерпретация хоть очень и напоминает GRW, но в одном важном пункте очень *отличается* от нее. Она постулирует коллапс не *всех* макротел, а лишь *конечного* наблюдателя, с точки зрения которого описывается физический эксперимент. При этом под словом «наблюдатель» мы понимаем не некую личность, обладающую загадочным «сознанием», а просто стандартный макроскопический объект, далекий от состояния термодинамического равновесия. Это совершено аналогично нашей предыдущей статье [1]. Для нас важно лишь то, что этот объект последний в цепочке наблюдений и его собственная термодинамическая стрела времени выбрана как «положительно» направленная. Это ограничение на коллапс приводит к серьезным последствиям, которых не было в GRW. А именно, если наличие коллапса в GRW теоретически можно проверить экспериментально, то проверить наличие коллапса в Копенгагенской Интерпретации невозможно даже *в принципе*. Обоснуем эту точку зрения. Рассмотрим мысленные эксперименты, которые позволяют проверить наличие коллапса, предсказываемого в GRW. Далее мы проверим, могут они же использоваться для проверки наличие коллапса в Копенгагенской Интерпретации.

a. Квантовая механика, как и классическая, предсказывает возвраты Пуанкаре. Причем, в отличие от классических хаотических систем, они происходят периодически или почти периодически. Наличие коллапса в GRW делает такие возвраты невозможными и не наблюдаемыми, т.е. этот факт можно проверить экспериментом
b. Квантовая механика обратима. При обращении эволюции система вернется в исходное состояние. Однако при коллапсе обратимость теряется. Этот факт также можно проверить экспериментом
c. Мы можем регистрировать экспериментально малые эффекты, связанные с малыми квантовыми корреляциями, которые остаются даже после декогеренции. В GRW эти малые эффекты исчезают.

Мы проверяем коллапс наблюдателя в Копенгагенской Интерпретации, и, следовательно, неизбежно должны включить его в наблюдаемую систему. Т.е. здесь происходит *самонаблюдение*. Покажем, что это не позволяет проверить (или опровергнуть) наличия коллапса в Копенгагенской Интерпретации описанными выше методами.

a. Предположим, мы будем ждать, когда наступит возврат, предсказываемый квантовой механикой. Но наблюдатель, коллапс которого мы проверяем, является неотъемлемой частью системы. Т.е. при возврате Пуанкаре, он вернется также в исходное состояние. Следовательно, вся его память о прошлом сотрется. Что сделает экспериментальную проверку, связанную со сравнением начального и конечного состояния не возможной.
b. Те же самые причины, что и в предыдущем пункте, сделают невозможной проверку возврата при обращении движения.
c. Для регистрации экспериментально малых эффектов, необходима очень высокая точность измерения. Но, поскольку наблюдатель сам входит в систему измерения (самонаблюдение) он не способен точно и полно измерить все параметры такой системы. Образно говоря, наблюдатель должен «записать» текущее состояние системы теми же самыми «чернилами», которые он, в том числе (как часть системы), и должен описать! На практике возможно описание лишь макропараметров системы, что делает невозможной наблюдение и расчет малых эффектов.

По сути дела, первые два пункта (a, b) связаны со следующим фактом, который имел место и в классической механике [1]. Декогеренция (расщепление на макросостояния) приводит к росту энтропии (одно макросостояние заменяется на целый набор возможных макросостояний). С другой стороны, наблюдение возврата (т.е. рекогеренция) связано с уменьшением энтропии. Наблюдатель же способен экспериментально проводить самонаблюдение лишь на ограниченных промежутках времени, когда у него существует стрела времени (т.е. состояние далекое от термодинамического равновесия), и она *не меняет* свое направление. Таким образом, невозможность экспериментально различить Копенгагенскую и Многомировую Интерпретации тесно связана с законом роста энтропии и термодинамической стрелой времени.

Все, сказанное выше, делает невозможной экспериментальную проверку разницы между Копенгагенской и Многомировой Интерпретацией, что делает их равноправными. Такие утверждения о неразличимости этих интерпретаций не раз встречаются в литературе. Однако там, где этот факт не только формулируют, но и пытаются доказать, обычно ссылаются на невозможность сделать такую проверку лишь *практически* для макротел, не понимая ее *принципиальной* невозможности. На этом основании делается ошибочный вывод о «привилегированном» положении Многомировой Интерпретации. Приведем наиболее наглядный пример [9]:

**"Сторонники Многомировой интерпретации могли бы утверждать, что, фактически, бремя экспериментального доказательства находится на противниках Многомировой интерпретации, потому что это - они, те, кто утверждает, что есть новая физика вне хорошо проверенного уравнения Шредингера".**
"Несмотря на название "интерпретация", Многомировой интерпретации - это просто вариант квантовой теории, которая отличается от других. Экспериментально, разность существует относительно теорий с коллапсом. Кажется, что нет никакого эксперимента, отличающего Многомировую интерпретацию от других теорий без коллапса, таких как механика Бома или другие варианты Многомировой интерпретации. Коллапс приводит к эффектам, которые являются, в принципе, наблюдаемыми; эти эффекты не существуют, если Многомировая интерпретация - правильная теория. Чтобы наблюдать коллапс, мы нуждались бы в технологии высшего качества, которая позволяет "обращать" квантовый эксперимент, включая инверсию процесса детектирования макроскопическими устройствами. См. Lockwood 1989 (p. 223), Vaidman 1998 (p. 257), и другие предложения в Deutsch 1986. Эти предложения - все для чисто мысленных экспериментов, которые не могут быть выполнены сейчас или с помощью какой-либо будущей технологией, достижимой в обозримое время. Действительно, в этих экспериментах должна наблюдаться интерференция различных миров. Миры различны, когда, по крайней мере, один макроскопический объект находится в макроскопически различимых состояниях. Таким образом, необходим интерференционный эксперимент с макроскопическим телом. Сегодня есть интерференционные эксперименты с большими и большими объектами (например, молекулы фуллерена $C_{60}$), но эти объекты все еще не являются достаточно большими, чтобы считаться "макроскопическими". Такие эксперименты могут только улучшить оценку границу, где коллапс мог бы иметь место. Решающий эксперимент должен включать интерференцию состояний, которые обладают макроскопическим числом степеней свободы: невозможная задача для сегодняшней технологии"

Приведенное здесь доказательство экспериментальной непроверяемости коллапса в Копенгагенской Интерпретации, насколько нам известно, встречается лишь в этой и предшествующих ей работах [10-13]. Его можно назвать «Геделевской» теоремой о невозможности для квантовой механики. Она как формулировкой, так и методом доказательства действительно напоминает «Геделевскую теорему о неполноте».

Мы здесь столь подробно останавливаемся на этом вопросе, поскольку Во-первых, сама невозможность экспериментально различить Копенгагенскую и Многомировую Интерпретации тесно связана с законом роста энтропии и термодинамической стрелой времени. Во-вторых, слишком много людей искренне, но ошибочно верят, что Многомировая Интерпретация (или иные менее модные Интерпретации) полностью решают все проблемы квантовой механики. К этим проблемам в первую очередь относятся уже сформулированные выше проблемы *неопределенности* в предсказаниях квантовой механики, не описываемые правилами Бора. Как же они решаются на самом деле? Это объясняется тем, что указанная неопределенность хоть и существуют в *идеальной* динамике, в *наблюдаемой* динамике отсутствует и экспериментально не наблюдаемо в *принципе*.

1) Самонаблюдение. Те же самые причины, уже описанные выше, которые не позволяют проверить коллапс экспериментально не позволят обнаружить экспериментально неопределенность, указанную в пунктах 1 (точный момент времени коллапса) и 2 (квантовые корреляции). А значит обсуждать ее бессмысленно.
2) Внешнее наблюдение.

a. Если это наблюдение не возмущает наблюдаемую систему, то коллапса системы а, следовательно, и неопределенности (указанной в пунктах 1 и 2) не возникает и квантовая механика может быть проверена точно экспериментально. Такое непертрубативное наблюдение возможно для макротел лишь теоретически, и только при условии известного начального состояния, чистого или смешанного. (Приложение A)

b. Присутствует малое взаимодействие между наблюдаемой системой и наблюдателем/окружением. Это малое взаимодействие маскирует неопределенность (указанную в пунктах 1 и 2) и делает невозможным ее экспериментальное наблюдение.

Здесь нужно вернуться к неопределенности, описываемой в пункте 3. Большинство реальных наблюдений отвечает случаям самонаблюдения (когда полное описание невозможно в принципе) или открытой системе, возмущаемой неконтролируемым малым внешним шумом от наблюдателя/окружения. Как же описывать такие открытые или неполные системы? Это делается путем ввода *макропараметров* системы. Реальная *наблюдаемая* динамика таких параметров возможна для широкого класса систем. Она не включает ненаблюдаемые в реальности «параллельные миры», уменьшение энтропии, квантовую суперпозицию макросостояний и другую экзотику, возможную только в *идеальной* динамике. Наблюдаемая динамика производится относительно термодинамической стрелы времени реального макроскопического неравновесного наблюдателя, слабо взаимодействующем с наблюдаемой системой и окружением (декогеренция). Идеальная динамика строится в абстрактном, координатном времени. Проблема перехода от идеальной к реальной динамике успешно решена в других работах [14-15, 17-18]. Выбор макропеременных неоднозначен, но и не произволен. Макропеременные должны выбираться так, чтобы при росте энтропии случайный малый внешний шум не влиял значительно на их динамику. Такие макропеременные существуют и называются главные переменные (pointer states) [3,17]. Наличие избранных состояний объясняется локальностью взаимодействия в реальном мире. Сильнее взаимодействую близкие частицы. Если бы сила взаимодействия определялась, например, близостью импульсов, то главные состояния были бы совсем иные. Так, поскольку свойство локальности неверно на расстоянии сравнимом с длинной волны, радиоволны имеют полевые pointer states, сильно отличающиеся от pointer states частиц. Ситуация, здесь описываемая, полностью эквивалентная [1], где рассматривались «подходящие» макросотояния для случая классической механики.

Что может служить примером наблюдаемой динамики для квантовых систем? Это указанные выше GRW теории. Для того чтобы понять это вернемся к Копенгагенской Интерпретации. Мы можем выбрать за «наблюдателя» в Копенгагенской Интерпретации разные неравновесные макротела. Теоретически, при этом коллапс будет видеться по-разному для таких разных наблюдателей. Это явление называется «парадокс друга Вигнера». Иначе это явление неоднозначности коллапса в Копенгагенской Интерпретации можно назвать «Квантовый солипсизм», по аналогии с похожим по смыслу философским учением. Разрешается он аналогично тому, как мы это делали в работе [1]. При синхронизации стрел времени энтропия всех слабо взаимодействующих тел растет (убывает) синхронно. Коллапс же соответствует именно росту энтропии (одно макросостояние заменяется на целый набор возможных макросостояний). Следовательно, малое взаимодействие (декогеренция) между макротелами приводит не только к синхронизации стрел времени, но и к синхронизации момента «коллапса» для разных наблюдателей. Это делает «Квантовый солипсизм» для макротел, хотя теоретически возможным, но крайне трудно осуществимым на практике. В этом заключается его отличие от Копенгагенской Интерпретации, где коллапс наблюдателя нельзя предотвратить даже теоретически. Указанные выше GRW теории являются, таким образом, для квантовой механики описанием действительной *наблюдаемой* динамики макротел (динамика FAPP). Она отбрасывают не наблюдаемые в реальности рассогласования коллапса макротел и убывание энтропии, которые предсказываются *идеальной* динамикой.

Хорошей иллюстрацией вышеописанной связи наблюдаемой и идеальной динамик является «парадокс котелка, который никогда не закипит». Он связан в квантовой механике с отклонением от экспоненциального вида закона распада частиц (или перехода с одного энергетического уровня на другой). Экспоненциальный характер такого закона очень важен – относительная скорость распада не зависит от момента времени. Это значит, что распадающаяся частица не имеет «возраста». В квантовой механике, однако, на малых временах закон *идеальной* динамики распада сильно отличается от экспоненциального. Это приводит к тому, что когда число измерений состояния частицы на ограниченном интервале времени увеличивается, частица в пределе вообще не распадается! Пусть мы наблюдаем макросистему, состоящую из большого числа распадающихся частиц. Здесь следует отметить, что распад частицы происходит по законам идеальной динамики лишь между измерениями. Сами измерения сильно влияют на динамику системы, как видно из формулировки самого парадокса. Чтобы перейти к *наблюдаемой* динамике, описанной выше, мы должны сильно уменьшить возмущающее влияние наблюдения. Это достигается увеличением интервала между наблюдениями, сравнимого со средним временем жизни отдельной частицы. Для таких больших интервалов времени, мы получаем реальную наблюдаемую динамику распада, при которой она описывается экспонентой, и среднее время жизни не зависит от конкретного интервала между измерениями. Таким образом, экспоненциальный распад – закон наблюдаемой, а не идеальной динамики частиц. (Этим же объясняется отсутствие возвратов Пуанкаре для такой системы.)

## 3. Количественное рассмотрение вопроса.
### 3.1 Определение основных понятий.

2) В классической механике микросостояние – точка в фазовом пространстве. В квантовой механике этому соответствует волновая функция $\psi$ (чистое состояние), а траектории – эволюция волновой функции во времени. В классической механике макросостояние соответствует функции распределея в фазовом пространстве. В квантовой механике этому соответствует матрица плотности $\rho$. Вид матрицы плотности зависит от выбранного базиса ортонормальных волновых функций. Если $\rho\rho \neq \rho$ - это смешанное состояния.

3) Уравнение движения для матрицы плотности $\rho$ имеет форму:

$$i\frac{\partial \rho_N}{\partial t} = L\rho_N,$$

где $L$ является линейным оператором:

$L\rho = H\rho - \rho H = [H,\rho]$

и H является оператором энергии для системы,

$N$ – число частиц

4) Если $A$ является оператором некоторой наблюдаемой, то средняя величина наблюдаемой может быть найдена следующим образом:

$<A> = tr\, A\rho$

5) Если производится самонаблюдение полный набор наблюдений сделать невозможно. В случае внешнего наблюдения из-за малого взаимодействия с наблюдателем и неустойчивости хаотической наблюдаемой системы полное описание также бессмысленно. Поэтому вводят ограниченный набор M *макропеременных*:

$A_{set} = \{A_1, A_2, ..., A_M\}$,

где $M << N$

Эти макропеременные известны с конечной малой погрешностью:

$\Delta A_i << A_i,\ 1 \leq i \leq M$

Этому набору макропеременных отвечает макросостояние с матрицей плотности $\rho_{set}$. При этом все микросостояния отвечающие условиям

$\{\ |<A_1> - A_1| \leq \Delta A_1,\ |<A_2> - A_2| \leq \Delta A_2,\ ...,\ |<A_M> - A_M| \leq \Delta A_M\}$

являются равновероятными.

*Термодинамическому равновесию* отвечает макросотояние $\rho_E$. Ему соответствует набор микросостояний, удовлетворяющий условию

$|<E> - E| \leq \Delta E\ (\Delta E << E)$,

где $E$ полная энергия системы.

При этом все эти микросотояния равновероятны.

6) В квантовой механике *энтропия ансамбля* определяется через матрицу плотности **[15]**:

$S = -k\, tr(\rho\, \ln \rho)$,

где $tr$ обозначает след матрицы

Энтропия, определенная таким путем, не меняется в процессе обратимой эволюции:

$\frac{\partial S}{\partial t} = 0$

7) *Макроскопическая энтропия* определяется следующим образом:

 a) Для данного $\rho$ находим все наборы макропеременных, ему соответствующих

$$\begin{cases} A_{set}^{(1)} = \{A_1^{(1)}, A_2^{(1)}..., A_M^{(1)}\} \Delta A_i^{(1)} << A_i^{(1)}, 1 \le i \le M \\ \vdots \\ A_{set}^{(L)} = \{A_1^{(L)}, A_2^{(L)}..., A_M^{(L)}\} \Delta A_i^{(L)} << A_i^{(L)}, 1 \le i \le M \end{cases}$$

b) Находим матрицу $\rho_{set}$ для которой все микросотояния, имеющие указанный набор макропараметров, равновероятны

c) Макроскопическая энтропия $S=-k\, tr(\rho_{set}\, ln\, \rho_{set})$

В отличие от ensemble entropy макроскопическая энтропия (макроэнтропия) не постоянна и может как увеличиваться, так и уменьшаться во времени. Она для данной энергии $E \pm \Delta E$ достигает максимума при термодинамическом равновесии. Направление роста макроэнтропии определяет направление собственной *термодинамической стрелы времени* системы.

8) Подобно классическому случаю локальность взаимодействия приводит к тому, что не все макросостояния приемлемы. Они должны быть выбраны так, чтобы малый шум не влиял существенно на эволюцию системы в направлении *термодинамической стрелы времени* системы. Подобные состояния хорошо исследованы и называются *pointer states* [3,17]. Квантовая суперпозиция таких состояний неустойчива относительно малого шума и не является, соответственно pointer state. Для макросистем, близких к состоянию термодинамического равновесия, в качестве pointer states удобно использовать собственные функции гамильтониана.

9) Огрубленная величина $\rho$ должна быть использована, чтобы получить изменение энтропии подобное макроскопической энтропии. Приведем пути, которыми этого можно добиться:

a) Мы определяем некий набор pointer states и проектируем матрицу плотности $\rho$ на эти состояния, т.е. отбрасываем недиагональные члены матрицы плотности $\rho$, записанной в представлении собственных функций этих наблюдаемых

$S=-k\, tr(\rho_{coar}\, ln\, \rho_{coar})$

b) Мы разбиваем систему на несколько взаимодействующих подсистем (например: наблюдатель, наблюдаемая система, окружение) и записываем полную энтропию как сумму энтропий этих подсистем:

$S=S_{ob}+S_{ob\_sys}+S_{env}$

## 3.2 Эффект слабого взаимодействия

### 3.2.1 Малое внешнее возмущение.

Мы можем окружить нашу макросистему ограниченного объема другой системой бесконечного объема («окружение», «резервуар»). Положим, что она находится в термодинамическом равновесии, имеет ту же среднюю температуру, что и наша система, и слабо взаимодействует с нашей ограниченной системой. Затем можно использовать квантовую версию «новой динамики», разработанной Пригожиным [14] для таких бесконечных систем. Полученная таким образом динамика нашей ограниченной системы (подсистема этой большой системы) будет совпадать с ее *наблюдаемой динамикой* в системе её собственного термодинамического времени, но без резервуара. Такое описание имеет смысл только в течение времени, когда стрела собственного термодинамического времени существует (т.е. система не находится в состоянии термодинамического равновесия) и не меняет своего направления.

### 3.2.2 Синхронизация термодинамических стрел времени при взаимодействии макросистем (наблюдателя и наблюдаемой системы).

Следует отметить, что здесь наша работа намного проще, чем в случае классической механики. Это связано с тем, что количественная теория малого взаимодействия между квантовыми системами (*декогеренция, запутывание*) – хорошо развитая область [2-3,17,24-27].
Мы не будем повторять его здесь, а подведем лишь краткие итоги.

   **(a)** Предположим, что мы имеем в некоторый момент времени две макросистемы, при этом одна из них или обе эти макросистемы находятся в квантовой суперпозиции их pointer states. Теория декогеренции [2-3,17,24-27] утверждает, что малое взаимодействие между макросистемами очень быстро (время декогеренции намного меньше времени релаксации к термодинамическому равновесию) переводит такую систему в смешанное состояние, при котором данная квантовая суперпозиция исчезает. Такой процесс исчезновения квантовой суперпозиции pointer states соответствует росту энтропии. Из теоремы Пуанкаре следует, что система (в координатном времени) должна вернуться в исходное состояние и должен произойти обратный процесс рекогеренции. Но он произойдет в обеих системах синхронно. Это означает, что в системе собственного термодинамического времени каждая из подсистем будет видеть лишь декогеренцию и рост энтропии. Это означает, что как процессы декогеренции, так и стрелы времени, будут синхронны во взаимодействующих подсистемах. Следует особо отметить, что мы рассматриваем здесь случай *макроскопических* систем. Для малых систем, где возможны большие флюктуации параметров, подобная синхронизация стрел времени и времен «коллапсов» подсистем не наблюдается [6-7].

   **(b)** Теперь предположим, что все макроскопические подсистемы находятся в их pointer states. В теории декогеренции показывается, что при наличии малого шума между ее макроскопическими подсистемами поведение квантовой системы полностью эквивалентно и неотличимо от поведения классической системы [2-3,17,24-27]. Таким образом, анализ синхронизации стрел времени здесь будет полностью эквивалентен, сделанному в работе [1].

   **(c)** Следует уточнить, что здесь понимается по словам «классическая» система.
Это означает, что в теории отсутствуют специфические математические особенности квантовой теории, такие как не коммутирующие наблюдаемые. При этом эти «классические» теории могут быть весьма экзотическими, включать в состав своих констант постоянную Планка и не сводится к законам обычной механики макротел.

Сверхпроводимость, сверхтекучесть, излучение абсолютно черного тела, опыты Фридмана с суперпозицией токов [19] часто называют «квантовыми». Они действительно квантовые в том смысле, что их уравнения движения включают постоянную Планка. Но они прекрасно описываются в макромасштабе математическим аппаратом обычных классических теорий: либо теорией *классического* поля (в качестве pointer states), либо теорией *классических* частиц (в качестве pointer states). С этой точки зрения, они не квантовые, а классические. В квантовой теории описываемые объекты являются частицами и волнами (вероятности) – одновременно.

Следует отметить, что в классическом пределе при комнатных температурах квантовая механика *массивных* частиц дает теорию *классических* частиц в качестве pointer states (пучки электронов, например), а для *легких* частиц в пределе получается *классическое* поле в качестве pointer states (радиоволны). Причем эти теории не включают постоянную Планка.

Однако, при высоких температурах вещества, когда излучение может идти уже на высоких частотах, световые кванты описываются теорией *классических* частиц в качестве pointer states и дают, например, спектр излучения абсолютно черного тела на высоких частотах. Хотя этот спектр включает постоянную Планка динамика его pointer states (частиц) будет классической. Для получения этого спектра формализм квантовой механики не нужен (Сам Планк этот спектр получил, ничего не ведая о математическом аппарате квантовой физики).

Наоборот, при низких температурах частицы начинают описываться *классическими* полями в качестве pointer states (явления сверхтекучести или сверхпроводимости). Например, сверхпроводимость описывается *классической волной* «параметра порядка». И хотя уравнения, описывающее это поле, включают постоянную Планка, но сами уравнения соответствуют математическому аппарату *классической* теории поля. Эти волны могут складываться между собой, аналогично квантовым. Но их квадрат амплитуды дают не плотность вероятности, а плотность куперовских пар. Такая волна не может коллапсировать при измерении, подобно вероятностной квантовой волне [20].

Для квантово-механических состояний бозонов при низких температурах pointer states отвечают *классические* поля, а при высоких температурах - *классические* частицы. Под словом «классический» понимается именно математический аппарат наблюдаемой динамики, описывающей их поведение, а не наличие или отсутствие постоянной Планка в уравнениях движения.

Что происходит в промежуточных состояниях между классическими полями и классическими частицами? Это, например свет в оптическом волноводе ($L \gg \lambda \gg \lambda_{ultraviolet}$), $L_{opt}$ – характерный размер макросистемы (оптического волновода) (Приложение B), $\lambda$- длина волны света, $\lambda_{ultraviolet}$ –ультрафиолетовая граница света). При использовании макромасштабов и макропеременных, а также с учетом малого шума от наблюдателя оба описания («классическая волна», «классический поток частиц») дают одинаковый результат и эквивалентны и могут быть использованы в качестве pointer states. Эквивалентная ситуация возникает для случая сверхпроводника, где роль частиц или волн играют элементарные «возбуждения» в газе куперовских пар.

Проведем простой расчет, иллюстрирующий вышесказанное.

Пусть E-энергия частицы; k –постоянная Больцмана, T –температура, p – импульс, $\lambda$ – длина волны частицы, $\omega$ –частота, $\Delta x$ – неточность координаты; $\Delta p$ –неточность импульса, $\hbar$ – постоянная Планка. Рассмотрим «газ» таких частиц, находящийся в полости, заполненного неким материалом с расстоянием между атомами a. a<<L, L- характерный размер полости. В вакууме $a \sim (L^3/N)^{1/3}$, N –число частиц в полости. c – скорость света (пусть для простоты показатель преломления вещества в полости близок к 1).

1) Возьмем вначале легкие частицы, которые при комнатной температуре имеют скорость, близкую к скорости света c.

$E \sim pc$; $E \sim kT$; $p \sim \Delta p$; $\lambda \sim \Delta x$; $\Delta p \Delta x \sim \hbar$; $\omega = E/\hbar$

Отсюда

$\hbar \sim \Delta p \Delta x \sim p \lambda \sim kT\lambda/c \Rightarrow \lambda \sim \hbar c/kT$

Условие *классического* полевого приближения с частотой $\omega \sim c/\lambda$:

$L < \lambda$ или $L \sim \lambda$. Отсюда $L < \hbar c/kT$ или $L \sim \hbar c/kT$

Условие приближения *классических* релятивистских частиц с $E \sim \hbar c/\lambda$ и $p=E/c$:

$L >> \lambda$. Отсюда $L >> \hbar c/kT$

2) Возьмем теперь тяжелые частицы *бозоны*, которые при комнатной температуре имеют скорость v<<c

$p \sim (Em)^{1/2}$; $E \sim kT$; $p \sim \Delta p$; $\lambda \sim \Delta x$; $\Delta p \Delta x \sim \hbar$; $\omega = E/\hbar$

Отсюда

$\hbar \sim \Delta p \Delta x \sim p \lambda \sim (kTm)^{1/2} \lambda \Rightarrow \lambda \sim \hbar/(kTm)^{1/2}$

Условие *классического* полевого приближения с частотой $\omega = p^2/(m\hbar)$:

$L < \lambda$ или $L \sim \lambda$. Отсюда $L < \hbar/(kTm)^{1/2}$ или $L \sim \hbar/(kTm)^{1/2}$

Условие приближения *классических* частиц с энергией $E=p^2/(2m)$ и импульсом $p=mv$:

$L >> \lambda$. Отсюда $L >> \hbar/(kTm)^{1/2}$

3) Возьмем теперь тяжелые частицы *фермионы*, которые при комнатной температуре имеют скорость v<<c

$p \sim (Em)^{1/2}$; $E \sim kT$; $p \sim \Delta p$; $\Delta p \Delta x \sim \hbar$

$\Delta x \leq \lambda$ и $\lambda \leq a$ – это условие связано с принципом Паули для фермионов. Они не могут оказываться в одном состоянии, каждый сидит в своей «коробочке» a.

Отсюда

$\hbar \sim \Delta p \Delta x \leq p \lambda \sim (kTm)^{1/2} \lambda \Rightarrow a \geq \lambda \geq \hbar/(kTm)^{1/2}$

$T \geq T_F = \hbar^2/(a^2 km)$ – температура Ферми, когда фермионный газ переходит в основное состояние и выражение $E \sim kT$ становится неверным.

При $T<T_F$: $E \sim E_F = kT_F$; $\lambda \sim \hbar/(E_F m)^{1/2} \sim a$

Условие *классического* полевого приближения:

$L < \lambda$ или $L \sim \lambda$. Но это невозможно, потому что $L >> a \geq \lambda$

При $T \geq T_F$ pointer states – это *классические* частицы с энергией $E=p^2/(2m)$ и импульсом $p=mv$.

При T≤$T_F$ pointer states – это *классические* частицы, заключенных в «ящички» размером a, с энергией E~$E_F$ и импульсом p~$(E_F m)^{1/2}$.

При T~$T_F$ мы наблюдаем динамику «возбуждений» в вырожденном Ферми-газе, которая описывается частицами или волнами в качестве pointer states для этих «возбуждений».

Чтобы создать в эксперименте ситуацию парадокса «Шредингеровского кота», нам необходимо суперпозиция именно pointer states, а не классических волн. Посему суперпозиция классических волн «параметра порядка» или световых волн никак не связана с этим парадоксом и не иллюстрирует его.

Так, например, в опытах Фридмана [19] состояния суперпозиции встречных токов сама является в данной ситуации pointer state. Это pointer state – классическая, а *не квантовая* суперпозиция pointer states, как неверно и стереотипно обычно считается. Действительно, состояние системы бозонов (куперовских пар) описывается при столь низких температура *классической* волной, как мы видели выше. Эти волны «параметра порядка» являются pointer states. Они отличаются от pointer states высокотемпературных электрических потоков *классических частиц*, имеющих определенное направление движения (тока). Суперпозиция, наблюдаемая в опытах Фридмана, не способна сколлапсировать в квантово-механическом смысле, поскольку ее квадрат описывает не вероятность, а плотность куперовских пар [20]. Она не более удивительная и не более «квантовая», чем обычная суперпозиция электромагнитных мод в закрытом резонаторе, где их спектр тоже дискретен. Единственное отличие - волновые уравнения «параметра порядка» в качестве pointer states включают ℏ. В этом и заключается вся их «квантовость».

### 3.3 Разрешение парадоксов Лошмидта и Пуанкаре в рамках квантовой механики.

Состояние квантовой хаотической системы в замкнутой полости с конечным объемом описывается набором энергетических мод $u_k(r_1, ..., r_N)$ со спектром $E_k$, распределенным по *случайному закону* [8].

Напишем уравнение для волновых функций невзаимодействующей пары таких систем:

$$\psi^{(1)}(r_1,...,r_N,t) = \sum_k u_k(r_1,...,r_N) e^{-\frac{iE_k^{(1)}}{\hbar}t}$$

$$\psi^{(2)}(r_1,...,r_L,t) = \sum_l v_l(r_1,...,r_L) e^{-\frac{iE_l^{(2)}}{\hbar}t}$$

Совместное уравнение следующее:

$$\psi^{(1)}(r_1,...,r_N,r_1,...,r_L,t) = \psi^{(1)}(r_1,...,r_N,t)\psi^{(2)}(r_1,...,r_L,t) =$$
$$\sum_k \sum_l u_k(r_1,...,r_N) v_l(r_1,...,r_L) e^{-\frac{i(E_k^{(1)}+E_l^{(2)})}{\hbar}t}$$

При наличии малом взаимодействия между системами

$$\psi^{(1)}(\boldsymbol{r}_1,...,\boldsymbol{r}_N,\boldsymbol{r}_1,...,\boldsymbol{r}_L,t) =$$

$$\sum_k \sum_l f_{kl}(\boldsymbol{r}_1,...,\boldsymbol{r}_N,\boldsymbol{r}_1,...,\boldsymbol{r}_L) e^{-\frac{iE_{kl}}{\hbar}t},$$

где $E_{kl} = E_k^{(1)} + E_l^{(2)} + \Omega_{kl}$, $\Omega_{kl}$-в общем случае набор случайных величин, $f_{kl}, u_k, v_l$ – собственные функции соответствующих Гамильтонианов.

Получающиеся решения являются почти периодическими функциями. Получающийся период возврата и определяет период Пуанкаре. Период возврата Пуанкаре совокупной системы является в общем случае больше периода обоих подсистем.

Для разрешения парадоксов Пуанкаре и Лошмидта (возвраты в этих парадоксах противоречат закону роста энтропии) рассмотрим теперь три случая

1) *Самонаблюдение*.

Поскольку при самонаблюдении собственная стрела времени всегда направлена по росту энтропии – то относительно этой стрелы времени наблюдатель способен видеть только рост энтропии. Кроме того, возврат в исходное состояние стирает всю память о прошлом, что не позволяет наблюдателю зафиксировать уменьшение энтропии. Таким образом, уменьшения энтропии и возвраты происходят лишь в координатном времени. В собственном времени наблюдателя (относительно которого и возможен любой эксперимент) они не могут быть экспериментально наблюдаемы [1,10-13].

2) *Внешнее наблюдение с малым взаимодействием* между макросистемами. Малое взаимодействие приводит к синхронизации стрел времени. Соответственно, все аргументы для самонаблюдения снова становятся релевантными.

3) Для очень трудноосуществимого *эксперимента с непертрубативным наблюдением* (Приложение А) уменьшение макроэнтропии действительно может наблюдаться. Однако, следует отметить, что в реальном мире энтропийные затраты на экспериментальную организацию такого непертрубативного наблюдения (наблюдаемую систему нужно очень сильно изолировать от шума окружения) намного превысят это уменьшение энтропии.

В классических системах период возврата Пуанкаре - это случайная величина, сильно меняющаяся в зависимости от начального состояния. В квантовых хаотических системах период точно определяется и не зависит значительно от начального состояния. Однако эта, казалось бы, реальная разница в поведении квантовых и классических систем не наблюдаема экспериментально даже в отсутствии ограничения на время эксперимента. Действительно, реальные физические эксперименты, возможно, проводить лишь на протяжении времени много меньшем периода Пуанкаре. Физические эксперименты имеют смысл только в течение времени, пока стрела собственного термодинамического времени существует (т.е. система не находится в состоянии термодинамического равновесия) и не меняет своего направления.

## 3.4 Декогеренция для процесса измерения.

### 3.4.1 Редукция системы при измерениях [22-23].

Рассмотрим ситуацию, когда прибор вначале находился в состоянии $|\alpha_0\rangle$, а объект — в суперпозиционном состояний $|\psi\rangle = \sum c_i |\psi_i\rangle$, где $|\psi_i\rangle$ — собственные состояния эксперимента. Начальный статистический оператор дается выражением

$$\rho_0 = |\psi\rangle |\alpha_0\rangle \langle\alpha_0| \langle\psi| \tag{1}$$

Парциальный след этого оператора, совпадающий со статистическим оператором системы, составленной из одного объекта, имеет вид

$tr_A(\rho_0)=\sum_n \langle\varphi_n|\rho_0|\varphi_n\rangle$

где $|\varphi_n\rangle$ — какая-то полная система состояний прибора. Таким образом,

$tr_A(\rho_0)=\sum |\psi\rangle\langle\varphi_n|\alpha_0\rangle\langle\alpha_0|\varphi_n\rangle\langle\psi|=|\psi\rangle\langle\psi|$, (2)

где использовано соотношение $\sum |\varphi_n\rangle\langle\varphi_n|=1$ и тот факт, что состояние $|\alpha_0\rangle$ нормировано. Мы получили в точности тот статистический оператор, который должны были приписать объекту, если бы он находился в состоянии $|\psi\rangle$. После акта измерения возникает корреляция между состояниями прибора и состояниями объекта, так что состояние комбинированной системы, составленной из прибора и объекта, описывается вектором состояния

$|\Psi\rangle=\sum c_i e^{i\theta i}|\psi_i\rangle|\alpha_i\rangle$. (3)

а статистический оператор дается выражением

$\rho=|\Psi\rangle\langle\Psi|=\sum c_i c_j^* e^{i(\theta i-\theta j)}|\psi_i\rangle|\alpha_i\rangle\langle\alpha_j|\langle\psi_j|$. (4)

Парциальный след этого оператора равен

$tr_A(\rho)=\sum_n \langle\varphi_n| \rho |\varphi_n\rangle =$
$=\sum_{(ij)} c_i c_j^* e^{i(\theta i-\theta j)} |\psi_i\rangle\{\sum_n \langle\varphi_n |\alpha_i\rangle\langle\alpha_j|\varphi_n\rangle\}\langle\psi_j|=$
$=\sum_{(ij)} c_i c_j^* \delta_{ij} |\psi_i\rangle\langle\psi_i|$ (5)

(так как различные состояния $|\alpha_i\rangle$ прибора ортогональны друг другу); таким образом,

$tr_A(\rho)=\sum |c_i|^2|\psi_i\rangle\langle\psi_i|$. (6)

Мы получили статистический оператор для системы, состоящей из одного объекта, описывающий ситуацию, когда имеются вероятности $|c_i|^2$ пребывать объекту в состояниях $|\psi_i\rangle$. Итак приходим к формулировке следующей теоремы.

**Теорема 5.5** (об измерении). Если две системы *S* и *A* взаимодействуют таким образом, что каждому состоянию $|\psi_i\rangle$ системы *S* соответствует определенное состояние $|\alpha_i\rangle$ системы *A,* то статистический оператор $tr_A(\rho)$ над полной системой (*S и A*) воспроизводит действие редукции, применяемого к акту измерения, производимого над системой *S*, находившейся до измерения в состоянии $|\psi\rangle=\sum_i c_i|\psi_i\rangle$. ∎

Метасостояние системы, находясь в котором она не имеет определенного состояния, но является частью большой системы, которая находится в чистом состоянии, называется *несобственным смешанным состоянием*.

### 3.4.2 Декогеренция при взаимодействии с макроскопическим прибором [21-23].

Учтем теперь, что прибор является макроскопической системой. Это означает, что каждая различимая конфигурация прибора (например, положение его стрелки) не является чистым квантовым состоянием, никоим образом ничего не утверждая о состоянии движения каждой отдельной молекулы стрелки. Таким образом, в вышеприведенном рассуждении начальное состояние прибора $|\alpha 0\rangle$ следует заменить некоторым статистическим распределением по микроскопическим квантовым состояниям $|\alpha_{0,s}\rangle$; начальный статистический оператор не дается выражением (1), а равен

$\rho_0 = \sum_s p_s | \psi \rangle| \alpha_{0, s}\rangle \langle\alpha_{0, s} |\langle \psi |$. (7)

Каждое состояние прибора $|\alpha_{0,s}\rangle$ будет реагировать на каждое собственное состояние $|\psi_i\rangle$ объекта тем, что превратится в некоторое другое состояние $|\alpha_{i,s}\rangle$, которое является одним из квантовых состояний, макроскопическое описание которого состоит в указании, что стрелка занимает положение i; точнее имеем формулу

$e^{iH\tau/\hbar}(| \psi \rangle| \alpha_{0, s}\rangle) = e^{i\theta_{i,s}}| \psi \rangle| \alpha_{i, s}\rangle$. (8)

Обратим внимание на появление фазового множителя, который зависит от индекса *s*. Разности энергий квантовых состояний $|\alpha_{0,s}\rangle$ с учетом времени τ должны быть такими, чтобы фазы $\theta_{i,s}(\mod 2\pi)$ были случайно распределены между 0 и $2\pi$.

Из формул (7) и (8) следует, что при $|\psi\rangle=\sum_i c_i|\psi_i\rangle$ статистический оператор после измерения будет даваться следующим выражением:

$$\rho = \sum_{(s, i, j)} p_s c_i c_j^* e^{i(\theta_{i,s} - \theta_{j,s})} |\psi_i\rangle|\alpha_{i,s}\rangle\langle\alpha_{j,s}|\langle\psi_j| \qquad (9)$$

Так как из (9) получаем тот же результат (6), то видим, что статистический оператор (9) воспроизводит действие редукции, примененной к данному объекту. Он также практически воспроизводит действие редукции, примененной к одному прибору («практически» в том смысле, что речь идет о «макроскопической» наблюдаемой). Такая наблюдаемая не различает разные квантовые состояния прибора, соответствующие одному и тому же макроскопическому описанию, т. е. матричные элементы этой наблюдаемой между состояниями $|\psi_i\rangle|\alpha_{i,s}\rangle$ и $|\psi_j\rangle|\alpha_{j,s}\rangle$ не зависят от *r* и *s*. Среднее значение такой макроскопической наблюдаемой *A* равно

$$\text{tr}(\rho A) = \sum_{(s, i, j)} p_s c_i c_j^* e^{i(\theta_{i,s} - \theta_{j,s})} \langle\alpha_{j,s}|\langle\psi_j|A|\psi_i\rangle|\alpha_{i,s}\rangle =$$
$$= \sum_{(i, j)} c_i c_j^* a_{i,j} \sum_s p_s e^{i(\theta_{i,s} - \theta_{j,s})} \qquad (10)$$

Так как фазы $\theta_{i,s}$ распределены случайным образом, суммы по s обращаются в нуль при i≠j; следовательно,

$$\text{tr}(\rho A) = \sum |c_i|^2 a_{ii} = \text{tr}(\rho' A). \qquad (11)$$

где

$$\rho' = \sum |c_i|^2 p_s |\psi_i\rangle|\alpha_{i,s}\rangle\langle\alpha_{j,s}|\langle\psi_j| \qquad (12)$$

Получаем статистический оператор, который воспроизводит действие редукции прибора. Если стрелка прибора наблюдается в положении i, состояние прибора при некотором s будет $|\alpha_{i,s}\rangle$, причем вероятность того, что оно будет именно состоянием $|\alpha_{i,s}\rangle$, равна вероятности того, что до акта измерения было состояние $|\alpha_{i,s}\rangle$. Таким образом, приходим к формулировке следующей теоремы.

**Теорема О декогеренции макроскопического прибора**. Пусть квантовая система взаимодействует с макроскопическим прибором таким образом, что возникает хаотическое распределение фаз состояний прибора. Пусть ρ — статистический оператор прибора после измерения, рассчитанный с использованием уравнения Шредингера, а ρ' — статистический оператор, полученный в результате применения редукции к оператору ρ. Тогда невозможно произвести такой эксперимент с макроскопическим прибором, который зарегистрировал бы различие между ρ и ρ'.

Это так называемая теорема *Данери — Лойнжера — Проспери* **[22]**.
Для широкого класса приборов доказано, что хаотичность в распределении фаз, о которой идет речь в теореме 5.6, действительно имеет место, если устройство является макроскопическим, хаотическим и ее начальное состояние неравновесное. Хаотичность фаз в этом случае имеет своим источником случайность спектра энергий (собственных значений Гамильтониана) для квантовых хаотических систем [8].

Отметим, что (12), хоть и выполняется с высокой точностью, является приближенным по отношению к (9). Отсюда часто делают вывод, что приведенное доказательство является FAPP. Т.е. квантовые корреляции лишь трудно измерить практически, фактически они продолжают существовать, и, следовательно, *в принципе* они измеримы. Это, однако, совершенно неверно. Действительно, из теоремы Пуанкаре следует, что система не будет оставаться в смешанном состоянии (12), а должна вернуться в исходное состояние (7). Это является результатом этих самых малых поправок, которые не учтены в (12). Тем не менее, описываемая здесь система $|\alpha_{i,s}\rangle$ соответствует случаю *самонаблюдения*, и поэтому не способна наблюдать экспериментально эти возвраты *в принципе* (как мы показали выше в разделе о разрешении парадоксов Пуанкаре и Лошмидта). Следовательно, эффекты этих малых поправок существуют лишь на бумаге в координатном времени идеальной динамики, но *экспериментально* не наблюдаемы в собственном термодинамическом времени наблюдаемой динамики. Приведенная здесь логика показывает, что Daneri-Loinger-Prosperi theorem на самом деле ведет к разрешению парадокса редукции, доказывая невозможность экспериментально различить полную и неполную редукцию, а не является FAPP решением

Приводимая здесь логика также очень напоминает статью Maccone [4]. Это не удивительно, ведь переход от (7) к (12) соответствует увеличению числа микросостояний и росту энтропии. А переход из (12) в (7) соответствует уменьшению энтропии. Соответственно наше утверждение об экспериментальной ненаблюдаемости остаточной квантовой корреляции эквивалентно утверждению о ненаблюдаемости уменьшения энтропии и доказывается теми же методами, что и в [4]. На эту статью было выдвинуто возражение [6], на которое Maccone не смог дать разумный ответ [28]. Попробуем дать его сами.

Определим здесь необходимые величины и проблему.
Пусть A – наш прибор, а C – измеряемая квантовая система.
Первая величина, взаимная энтропия $S(A:C)$ – это огрубленная энтропия ансамбля, полученная разделением на две подсистемы, минус сама энтропия ансамбля. Поскольку вторая величина постоянна во времени эта величина хорошо описывает поведение макроэнтропии:

$S(A:C) = S(\rho_A) + S(\rho_C) - S(\rho_{AC})$,

где $S = -tr(\rho \ln \rho)$,
Вторая величина $I(A:C)$ - классическая взаимная информация определяет какую максимальную информацию о измеряемой системе ($F_j$) мы можем получить наблюдая показания прибора ($E_i$). Чем больше корреляция между системами, тем выше эта информация:

$I(A:C) = max_{E_i \otimes F_j} H(E_i:F_j)$, где

$H(E_i:F_j) = \Sigma_{ij} P_{ij} \log P_{ij} - \Sigma_i p_i \log p_i - \Sigma_j q_j \log q_j$

и $P_{ij} = Tr[E_i \otimes F_j \rho_{AC}]$, $p_i = \Sigma_j P_{ij}$ и $q_j = \Sigma_i P_{ij}$

данные POVMs (операторы наблюдаемых) $E_i$ и $F_j$ для A и C соответственно
Maccone [4] доказывает неравенство
$S(A:C) \geq I(A:C)$ (13)
И из него делает вывод, что убывание энтропии влечет за собой уменьшение информации (памяти) о системе. Но (13) является неравенством. Соответственно в [6] приводится пример квантовой системы из 3 кубитов, при которой убывание энтропии сопровождается ростом информации, хотя неравенство (13) продолжает выполняться.
Посмотрим, что происходит в нашем случае
До измерения (7)
$S(A:C) = -\Sigma_s p_s \log p_s + 0 + \Sigma_s p_s \log p_s = 0$

$E_i$ –соответствует набору $|\alpha_{0,s}\rangle$, $F_j$ - $|\psi\rangle$
$I(A:C) = -\Sigma_s p_s \log p_s + 0 + \Sigma_s p_s \log p_s = 0 = S(A:C)$
В конце измерения из (12)
$S(A:C) = -\Sigma_i |c_i|^2 \log |c_i|^2 - \Sigma_{s,i} |c_i|^2 p_s \log |c_i|^2 p_s + \Sigma_{s,i} |c_i|^2 p_s \log |c_i|^2 p_s = -\Sigma_i |c_i|^2 \log |c_i|^2$

$E_i$ –соответствует набору $|\alpha_{i,s}\rangle$ , $F_j$ - $|\psi_j\rangle$

$I(A:C) = -\sum_i |c_i|^2 \log |c_i|^2 - \sum_{s,i} |c_i|^2 p_s \log |c_i|^2 p_s + \sum_{s,i} |c_i|^2 p_s \log |c_i|^2 p_s = -\sum_i |c_i|^2 \log |c_i|^2 = S(A:C)$

Таким образом, наш случай соответствует

$$I(A:C) = S(A:C) \qquad (14)$$

в (13). Никаких проблем нет. Что и не удивительно – случай равенства в (13) соответствует именно макроскопической хаотической системе. Приводимая в возражении [6] система не является микроскопической. Это отражает тот широко известный факт, что такие понятия как термодинамическая стрела времени, возрастание энтропии и измерительный прибор относятся к макроскопическим хаотическим системам. Как сама статья [6], так и последующая за ней статья [7] описывают не термодинамическую стрелу времени, а сильно флуктуирующие небольшие системы, для которых никакая термодинамика невозможна. Полезным результатом этой работы можно считать равенство (14), которое может служить хорошим математическим критерием *критерий макроскопичности* хаотической системы, а разница между величинами в нём – мерилом ее флуктуаций.

Статья David Jennings, Terry Rudolph "Entanglement and the Thermodynamic Arrow of Time" очень интересна. Но понятие *Термодинамической* Стрелы Времени не применимо для микросистем. Это - хорошая статья о квантовых флуктуациях, но не статья о *Термодинамической* Стреле Времени. В Анонсе статьи "Entanglement and the Thermodynamic Arrow of Time" авторы пишут: "Мы исследуем подробно случай трех кубитов, и также предлагаем некоторые простые эксперименты, возможные с небольшим числом кубитов." Но никакая термодинамика не возможна для такой микросистемы. David Jennings, Terry Rudolph (как и Maccone) не понимают, что категория "*термодинамическая стрела времени*" применима только для больших макросистем. Использование этого понятия для небольшой флуктуирующей системы не имеет никакого физического смысла. Они также (как и Maccone) используют неправильное определение макроскопической *термодинамической* энтропии. Мы даем (вместо Maccone) правильный отклик на "Comment on "Quantum Solution to the Arrow-of-Time Dilemma"". *Правильный* отклик состоит в том, что никакие противоречия (найденный в этом Комментарии) не проявляются для макроскопических систем. Только для микроскопической системы такие противоречия существуют. Но понятия "Термодинамическая Стрела Времени" и "закон возрастания энтропи" не применимы для таких систем. Мы иллюстрируем этот факт рассмотрением квантовой хаотической макросистемы и демонстрируем, что противоречие (найденный David Jennings, Terry Rudolph для микроскопической системы) не существует для этого правильного термодинамического случая. Следует упомянуть, что большой размер системы (квантовой или классической) является не достаточным условием для системы, чтобы быть макроскопической. Макроскопическая система (рассматриваемая в Термодинамике) должна также быть хаотической (квантовой или классической) и иметь небольшое хаотическое взаимодействие с окружающей средой/наблюдателем, приводящее к декогеренции (для квантовой механики) или декорреляции (для классической механики). Нужно также упомянуть, что терминология, напоминающая термодинамическую терминологию, широко и эффективно используется в квантовой механике, квантовых компьютерах и теории информации. Большое число примеров может быть найдено в ссылках статьи Jennings и Rudolph. Другой хороший пример - энтропия Шеннона в теории информации. Но обычно авторы, использующий такую термодинамически-подобную терминологию, не рассматривают свою статью как анализ классической Термодинамики. Наоборот Jennings и Rudolph "опровергают" второй закон Термодинамики на основе нерелевантной микроскопической системы (в их Comment). Они делают (в этом же Comment) заявление об их следующей статье "Entanglement and the Thermodynamic Arrow of Time" как о правильном рассмотрении, опровергающем второй закон термодинамики.

## 4. Заключение.

В статье проводится анализ термодинамической стрелы времени для квантовых систем. Он во многом аналогичен классическому случаю. Важным отличием квантовых систем от классических является наличие микросостояний, которые соответствуют не одному макросостоянию, а целому их набору (квантовая суперпозиция макросотояний). Рассмотрение термодинамической стрелы времени для этого случая с помощью теории декогеренции дает разрешение парадоксов, связанных с редукцией (коллапсом) волнового пакета.

## Приложение А.   Непертрубативное наблюдение в квантовой и классической механике.

Часто можно столкнуться с утверждением, что в классической механике в принципе всегда можно организовать непертрубативное наблюдение**.** С  другой стороны в квантовой механике взаимодействие наблюдателя с наблюдаемой системой при измерении неизбежно. Покажем, что оба этих утверждения в общем случае неверны.

   Позвольте нам сначала определять невозмущающее наблюдение [10-11,30-31] в квантовой механике. Предположим, что у нас есть некоторая квантовая система  в известном начальном состоянии. Это начальное состояние может быть любой результатом некоторой подготовки (например, атом переходит в основание электронное состояние в вакууме в течение долгого времени) или результатом эксперимента по измерению (система QM после измерения может быть хорошо определенное состояние, соответствующее собственной функции измеренной переменной). Мы можем предсказать дальнейшую эволюцию начальной волновой функции. *В принципе* мы можем делать дальнейшие измерения, выбирая измеряемые переменные таким образом, чтобы соответствующие им наборы собственных функции в момент измерения  включали в себя текущую волновую функцию наблюдаемой системы. Такой измерительный процесс может позволить нам непрерывное наблюдение без любого возмущения наблюдаемой квантовой системы. Это невозмущенное наблюдение может быть легко обобщено для случая известного *смешанного* начального состояния. Действительно, в этом случае измеряемая переменная в каждый момент времени должна соответствовать такому набору собственных функций, в представлении которых матрица плотности в этот же момент времени будет диагональной.

   Например, позвольте нам рассматривать некоторый квантовый компьютер. У него есть некоторое четкое начальное состояние. Наблюдатель, которому известно это начальное состояние может *в принципе,* провести невозмущенное наблюдение любого промежуточного состояния квантового компьютера.

   Следует особо отметить, что подобное непертрубативное наблюдение возможно только при условии известного начального состояния. Но, наблюдатель, который не знает начальное состояние, не сможет сделать такое наблюдение, потому что он не может предсказать промежуточное состояние квантового компьютера**.**

   Рассмотрим теперь классическую механику. Пусть на вершине конуса лежит песчинка, *бесконечно* малого радиуса. Система находится в поле тяжести Земли. Тогда попытка пронаблюдать систему даже с *бесконечно малым возмущением*  приведет к нарушению равновесия с неопределенным будущим через *конечный* интервал времени. Конечно, приведенный пример экзотичен – он соответствует сингулярному потенциалу и бесконечно малому телу. Тем не менее, подобные сильно неустойчивые системы являются хорошими классическими аналогами квантовых систем. Среди них можно искать аналогии с квантовыми системами и квантовыми парадоксами. Введя условие, что классическое измерение оказывает очень малое, но не нулевое возмущение на измеряемую систему, можно снизить требования к сингулярности этих систем.

Очень часто приводят примеры «чисто квантовых парадоксов», якобы не имеющих аналогии в классической статистической механике. Одним из них является парадокс Элитцура-Вайдмана [29] с бомбой, которую можно обнаружить без взрыва:

*Пусть волновая функция одного кванта света разветвляется по двум каналам. В конце эти каналы снова объединяются, и происходит интерференция двух волн вероятности. Внесение в один из каналов бомбы нарушит процесс интерференции и позволит таким образом обнаружить бомбу, даже если квант света не подорвет ее, пройдя по другому каналу. ( Квант света считается способным взорвать бомбу)*

Классической аналогией этой ситуации является следующий эксперимент классической механики:

*В один из каналов, где нет бомбы, запустим макроскопический поток многих частиц. В другой канал, где, может быть, есть бомба, направим одновременно только одну **бесконечно** легкую частицу. Такая частица не способна взорвать бомбу, но она может быть отклонена ею назад. Если бомбы нет, то частица пройдет канал. На выходе этого канала с бомбой расположим описанный выше конус с песчинкой (**бесконечно** малого радиуса) на его вершине. Если наша бесконечно легкая части собьет песчинку с вершины, то это означает, что бомбы нет. Если песчинка останется на вершине после выхода потока частиц из второго канала, то это означает, что бомба есть.*

В данном примере бесконечно легкая частица является аналогом «невесомой» волновой функции квантовой частицы. Но квант света чувствителен к поведению этой «невесомой» волновой функции. Также и песчинка (бесконечно малого радиуса) на вершине конуса чувствительна по отношению к бесконечно легкой частице.

Подводя итог, можно сказать, что разница между квантовыми и классическими системами не столь принципиальна, как часто считается.

**Приложение В. Разложение на моды при произвольных граничных условиях**.

Часто возникает задача описание излучения в замкнутой полости, заполненной каким-либо веществом. Как правило, это делается путем разложения излучения на моды. Эти моды являются набором функций, на которые раскладывается любое излучение в некоторой полости и при некоторых граничных условиях. Например, это квадратная полость спериодическими граничными условиями. Затем полученное разложение подставляется в уравнение движения для излучения, где члены ряда почленно дифференцируются. Таким образом, получаются такие характеристики излучения, как $\omega(\mathbf{k})$, где $\omega$ – частота моды, а $\mathbf{k}$ – волновой вектор моды, $|k|=2\pi/\lambda$, $\lambda$ –длина волны моды.

Но тут возникает чисто математическая проблема. Для почленной дифференцируемости ряда требуется равномерная сходимость ряда во всех точках пространства. Это автоматически верно для любого излучения с такой же формой полости и граничными условиями, при которых были найдены моды. Но для любого другого случая это не так. Моды образуют полный ортогональный набор и любое излучение можно представить как суперпозицию таких мод. Но в общем случае ряд сходится неравномерно (плохо сходится около границ полости) и не может быть почленно продифференцирован. О проблеме несоответствия мод разложения и граничных условий пишет Peierls [32]. Однако он рассматривает случай, когда при данных граничных условиях существует некий полный ортонормальный набор мод, таким условиям удовлетворят. Но возможны ситуации, когда для данных граничных условий такого набора мод просто нет. Или нам не известны граничные условия, а заданы лишь энергетические условия на границе. Как же решается проблема в этом случае?

Дело в том, что все возмущения в излучении распространяются со скоростью, не превышающей скорость света в веществе полости $v=c$. Это значит, что любое возмущение в начальных условиях на поле излучения, возникшее в точке x, проявятся в точке $x_1$ только через конечное время $(x-x_1)/c$. Это значит, что возмущения от стенок достигнут центра полости за время $t=L/c$, где L - характерный размер полости. Неравновесная сходимость проявляется у ряда разложения излучения на моды только вблизи стенок полости. Внутри полости точная функция почти точно совпадает с рядом мод в течение времени L/c. Поэтому в этой области и в течение этого времени почленное дифференцирование даст почти точный результат и имеет смысл.

Чтобы верно оценит частоту моды ω(**k**) нужно, чтобы их амплитуда не менялась из-за возмущения от стенок существенно в течение времени много большего периода ее колебаний 2π/ ω(**k**) . Отсюда условие макроскопичности полости:

2π/ω<< L/c

или

L>>2π (c/ω)

ω – отвечает максимуму частот ω(**k**) .

Пусть условие макроскопичности полости выполняется.

Это значит, что почленное дифференцирование мод вдали от стенок полости дает верный результат на временных масштабах порядка 2π/ω.

На временных масштабах L/c результат не может быть верен. Здесь обычно используют соображения, основанные на законах сохранения энергии и роста энтропии. С помощью них и получается медленная эволюция амплитуд A(t, **r**) и фаз φ(t, **r**) мод:

E(t, **r**)=$\Sigma_i$ $A_i$(t, **r**)sin(ω($\mathbf{k}_i$)t+ $\mathbf{k}_i\mathbf{r}$+$\varphi_i$(t, **r**))

Для вакуума:

ω(**k**)=c|**k**|
L>>λ

## Благодарности

Я благодарю Hrvoje Nikolic и Vinko Zlatic за обсуждения и дискуссии, которые очень помогли при написании этой статьи.

## Библиография